\documentclass[twocolumns]{aastex631}

\shorttitle{Stars on the Boundary of the ECA and LCC}
\shortauthors{Varga et al.}

\graphicspath{{./}{Revised Figures/}}

\begin{document}

\title{Walking the Line: Young Stars on the Boundary of the Epsilon Cha and Lower Centaurus-Crux Associations}

\author[0000-0002-8301-4680]{Attila Varga}
\affiliation{College of Science, Rochester Institute of Technology, 54 Lomb Memorial Drive, Rochester, NY 14623, USA}

\author[0000-0002-3138-8250]{Joel H. Kastner}
\affiliation{Center for Imaging Science, Rochester Institute of Technology, 54 Lomb Memorial Drive, Rochester, NY 14623, USA
}
\author[0000-0002-4555-5144]{D. Annie Dickson-Vandervelde}
\affiliation{Department of Physics and Astronomy, Vassar College, 124 Raymond Avenue, Poughkeepsie, NY 12604, USA}

\author{Alex Binks}
\affiliation{Institut f\"ur Astronomie \& Astrophysik, Eberhard-Karls-Universit\"at T\"ubingen, Sand 1, 72076 T\"ubingen, Germany}

\begin{abstract}

Gaia Data Release 3 (DR3) has provided the largest and most astrometrically precise catalogue of nearby stars to date, allowing for a more complete membership census of nearby, young stellar moving groups. These loose associations of young (age $<$100 Myr) stars within $\sim$100 pc are vital laboratories for the study of the early evolution of low-mass stars and planetary systems. We have exploited DR3 data to examine the boundary region between two of the youngest nearby moving groups, the $\sim$3--8 Myr-old $\epsilon$ Cha Association (ECA) and an $\sim$8 Myr-old sub-population of the sprawling Lower Centaurus Crux (LCC) young star complex. Using spatio-kinematic and color-magnitude criteria designed to select stars in the ECA, we identify $\sim$54 new young-star candidates that extend from the ECA core to the southern edge of the LCC. Included among our new candidates are six previously unidentified ultra-low-mass, mid- to late-M stars, lying near the future hydrogen-burning limit, that display significant infrared excesses. Our spatial, kinematic, and CMD analysis of these new candidates and previously established LCC and ECA members blurs the boundary between these groups and provides evidence for a wave of continuous star formation extending from north (LCC) to south (ECA).  We discuss the factors which studies of nearby young moving groups must consider when constraining the ages of stars in these groups.

\end{abstract}

\section{Introduction} \label{sec:intro}

Identifying comoving associations of young (age $<$100 Myr) stars within $\sim$100 pc has been a focus of stellar astronomy for a few decades, as these nearby young groups are vital laboratories for the study of the early evolution of low-mass stars and planetary systems \citep{zuckerman_young_2004,torres_young_2008}. 

Recently, with advances in adaptive optics and coronagraphy, young nearby stellar associations have become especially important, providing prime targets for direct imaging and spectroscopy of young exoplanets 
\citep[e.g.,][]{desidera_sphere_2021,hinkley_jwst_2022,currie_direct_2023}. If a star can be confirmed to be a member of a given association, its relative age is immediately well constrained, and age estimates can then be further refined via isochronal-based methods and model atmosphere fitting. In the era of ESA's Gaia Space Astrometry Mission \citep{prusti_gaia_2016,gaia_collaboration_gaia_2023} astronomers now have the ability to measure the positions, space motions, and color-magnitude diagram (CMD) positions of the majority of stars in our solar neighborhood with unprecedented accuracy. The exploitation of Gaia data has allowed the identification of nearby young stars in moving groups to rapidly progress over the past half-dozen years, from a few hundred to thousands of members and candidate members \citep{gagne_banyan_2017,kerr_stars_2021}. 

The nearest OB association, the Scorpius-Centaurus Association (hereafter Sco-Cen, or SCO) \citep{zeeuw_hipparcos_1999,mamajek_post-t_2002}, serves as a prime example. Since its identification \citep[][and references therein]{mamajek_post-t_2002} 
it has been the focus of numerous studies identifying low mass members, exoplanet candidates, and debris disks, and this work has been greatly accelerated in the Gaia era  \citep{mann_zodiacal_2016,pecaut_star_2016,wright_kinematics_2018,rizzuto_tess_2020}. At an age of $\sim$10--20 Myr \citep{mamajek_post-t_2002,pecaut_star_2016} and with well in excess of 10000 members\citep{luhman_census_2021}, it provides one of the largest collections of young stars within $\sim$200 pc of Earth.

The overall structure of Sco-Cen has been widely studied, revealing several sub groups which can be differentiated from each other based on ages and Galactic space motions. One prominent subgroup is the Lower-Centarus Crux (LCC). With an age $\sim$5--20 Myr and distance $\sim$90--140 pc and suffering little intervening extinction \citep{pecaut_star_2016,luhman_census_2021}, this group has shown its promise, yielding two transiting exoplanet discoveries \citep{mann_zodiacal_2016,rizzuto_tess_2020,wood_tess_2023}, one directly imaged exoplanet \citep{lafreniere_directly_2010} and hundreds of disk-hosting young stars \citep{lieman-sifry_debris_2016}. Recent Gaia-based spatio-kinematic studies further uncovered even more substructure, revealing around five distinct subgroups \citep{goldman_large_2018,kerr_stars_2021}. 

Just beyond the southern end of the LCC lies the $\epsilon$ Cha Association (hereafter ECA). At an estimated age of $\sim$3--8 Myr, the ECA is the youngest of the many known young moving groups lying within $\sim$100 pc \citep[][hereafter M13 and DV+21]{murphy_re-examining_2013,dickson-vandervelde_gaia-based_2021}. Despite its youth, most ECA members suffer little interstellar extinction (M13, DV+21). The ECA has a relatively small spatial extent, with a radius $\sim$10 pc, while the LCC is $\sim$60 pc in extent. The centers of the two groups are separated by only $\sim$15 pc. Recent studies of the ECA have repeatedly discussed its potential relationship to the LCC \citep{murphy_re-examining_2013,kubiak_new_2021,kerr_stars_2021}. Indeed, in light of the spatial and kinematic proximity of the two groups, it has been proposed that the ECA is the most recently formed branch of the LCC \citep{kubiak_new_2021}. 

Given their young ages, we expect many members of both the ECA and LCC to be orbited by dusty circumstellar disks that are either actively forming planets (i.e., gas-rich protoplanetary disks) or constitute recent remnants of the planet formation process (i.e., gas-poor debris disks). 
Indeed, the ECA features star/disk systems, such as T Cha \citep[e.g.,][and references therein]{bajaj_jwst_2024} and MP Mus \citep[e.g.,][and references therein]{grimble_empirical_2024}, that represent among the nearest and hence most readily accessible examples of protoplanetary disks around young, Sun-like stars. 
More generally, the proximity of the ECA and LCC makes this region ripe for further statistical investigations of the likelihood of survival of dusty protoplanetary and debris disks to ages $\sim$5--15 Myr, via analysis of the frequency of detections of infrared excesses around bona fide and candidate members \citep{murphy_re-examining_2013,dickson-vandervelde_gaia-based_2021,pfalzner_low-mass_2024}. The ECA/LCC region is particularly promising for studies of dusty disks around extremely low-mass stars and (future) brown dwarfs \citep{luhman_census_2021,luhman_census_2021-1,pfalzner_most_2022}. 

In this paper, we probe the population of young stars that lie in the region of potential spatial and kinematic overlap between the ECA and LCC. We use the recent Gaia Data Release 3  \citep[DR3;][]{vallenari_gaia_2023} to identify potential new ECA members, and to analyze the positions and velocities of these candidate ECA stars alongside those of previously identified LCC members, so as to investigate the boundary between these groups. We derive fundamental parameters such as effective temperatures, luminosities, and estimated ages for our new candidates as well as previously identified young stars in this region, and we identify potential examples of infrared excesses among them.Section 2 describes our selection methodology for new ECA candidates. Section 3 presents analysis of the spatio-kinematics and age distribution of our new candidates, including comparisons with established ECA and LCC members, and highlights new candidates with infrared excesses. We discuss the results in Section 4, and summarize our conclusions in Section 5.

\section{New ECA Candidates}\label{sec:candidate overview section 2}
\subsection{Candidate Selection}\label{sec:selection}

\begin{figure}
    \centering
    \includegraphics[width=1.0\textwidth]{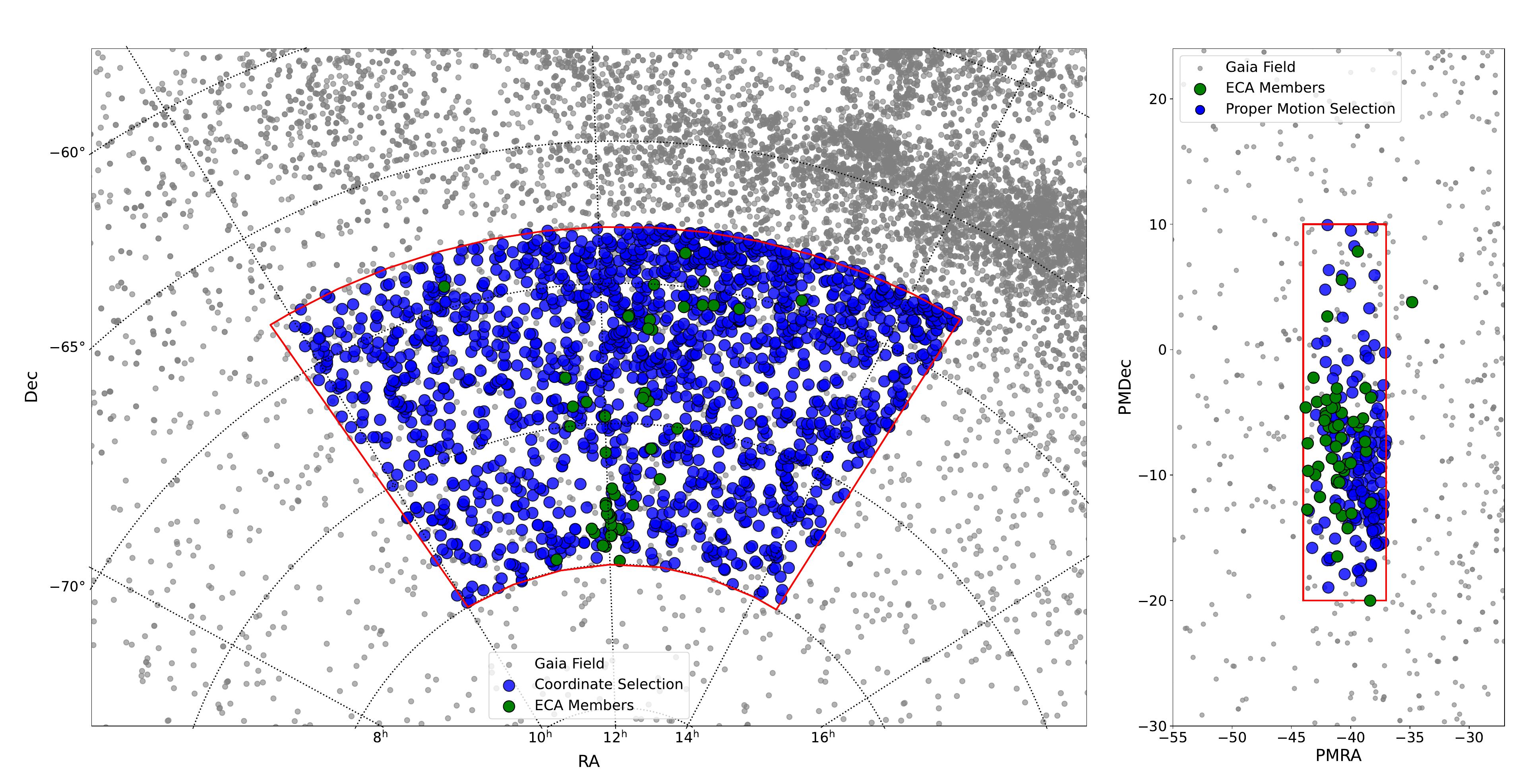}
    \caption{Left: Initial positional selection of our ECA candidates. Stars indicated in blue are those selected to lie within a 70 degree by 12 degree box centered on RA 180$^\circ$, DEC $-$74$^\circ$, and with parallaxes between $ 8.9 < \pi < 11$. The positions of previously identified ECA members and candidates are indicated in green. Right: Proper motion candidate selection. Filtered stars, again indicated in blue, are those with PMs between right ascension (PMRA) $-$44 and $-$37 mas yr$^{-1}$, and PMs in declination (PMDEC) between $-$20 and +10 mas yr$^{-1}$. Proper motions of previously identified ECA members and candidates are again indicated in green. In both panels, grey points indicate the background sample of stars from Gaia DR3 around and within our regions of interest.}
    \label{fig:pospm}
\end{figure}

Our investigation of the ECA/LCC boundary was initiated by retrieving potential ECA members that satisfy the DV+21 ECA membership criteria from Gaia DR3. 
We searched a 70 degree by 12 degree box centered on (RA, dec) 180$^\circ$, $-$74$^\circ$,  the coordinates of the approximate center of the ECA determined by DV+21, limiting the search to stars with parallaxes between $ 8.9 < \pi < 11$, i.e., the range listed for confirmed and candidate stars by DV+21 (Fig. \ref{fig:pospm}). 
This provides an initial list of 1643 stars. 
To refine this list, we follow \citet{riello_gaia_2021} and exclude stars with suspect photometry by eliminating those that fail to satisfy the criterion placed on Gaia's $\tt{photometric\_B_{p}\_R_{p}\_excess}$ ($E$) by their equation 6.

We further filter out stars with poorly defined 5-parameter astrometric solutions. We remove stars with large uncertainties on their measured parallaxes, i.e., stars with a parallax error $\pi_{err} > 0.1\pi$. We extend this same criterion to the astrometric excess noise quantity $\epsilon$, 
a hybrid noise parameter that considers a star's parallax and proper motion measurements, excluding stars with $\epsilon > 0.1\pi$. In certain cases this may be too weak a criterion. However, as outlined in the Gaia data release documentation (\url{https://gea.esac.esa.int/archive/documentation/GDR2/}), large astrometric excess noise may be due to binarity and/or systematic modeling errors. Our final quality check is to consider the renormalized astrometric unit weight error (RUWE), as it has been established that values of RUWE $>$ 1.4 may indicate unreliable photometric data \citep{lindegren_gaia_2021}. We do not apply any cutoff for this error but note the stars that have high RUWE. Among our final sample 22 out of 54 stars have an RUWE $>$ 1.4. We note that all these stars are faint M dwarfs close to Gaia's detection limit and therefore may exhibit high RUWE.
After removing stars on the basis of the foregoing Gaia data quality indicators, we consider stars in a reasonable parameter space in proper motion (PM), using a PM boundary encompassing the extrema of measured PMs reported by DV+21. Specifically, we include PMs in right ascension (PMRA) between $-$44 and $-$37 mas yr$^{-1}$, and PMs in declination (PMDEC) between $-$20 and +10 mas yr$^{-1}$, and filter out stars that lie outside this PM space.  
Next we ensure our candidates are consistent in isochronal age with the ECA. Stars of type K and M that lie more than $\sim$0.75 mag above the main locus of main sequence field stars in Gaia color-magnitude diagrams are likely to be young and/or magnetically active \citep[e.g.,][]{kastner_nearby_2017}. Here, we make use of the 5-8 Myr empirical isochrone derived by DV+21, based on their re-evaluation of the membership of the ECA, i.e., 
\begin{equation}
\label{eq:em iso}
G = 1.28c^4 + 8.44c^3 - 28.9c^2 + 33.17c - 7.18
\end{equation}
where $c=G - R_{p}$.

As outlined in DV+21, we define an absolute magnitude offset to discard stars that deviate from the empirical isochrone. This offset is given by
\begin{equation}
\label{eq:mag}
\Delta M = M'(c) - M
\end{equation}
where $M'(c)$ is the absolute magnitude a star of the same color would have if on the empirical isochrone. 
We remove stars from membership consideration if $|\Delta M| > 2\sigma_{\Delta M}$, where $\sigma_{\Delta M} = 0.629$ mag, the standard deviation in $\Delta M$ derived by DV+21 using the original members of the ECA listed in M13. This condition on $|\Delta M|$ effectively removes interloping field stars near the main sequence while avoiding eliminating binaries that lie above the empirical isochrone derived from the previously established ECA member stars included in DV+21.

We finalize our sample of new ECA candidates by consulting each star's reference and excluding stars previously identified as LCC or ECA members by \citet{goldman_large_2018} and \citet{murphy_re-examining_2013}. During this process some of our new ECA candidates were found to have references associated with the greater Sco-Cen complex \citep[e.g.,][]{song_new_2012,gagne_banyan_2018,zari_3d_2018}. However, these studies generally do not consider each star's position and space motion relative to the ECA and therefore we do not remove stars with these references. Most of our new ECA candidates also have references that indicate their likely youth \citep[e.g.,][]{gagne_banyan_2018,kerr_stars_2021,zari_3d_2018}, supporting our identification of these stars as being photometrically young and potential members of the ECA. The resulting final sample of 54 previously unidentified ECA candidates is listed in Table \ref{table:gaia param}, with 17 of these stars having no previous reference in the literature.

\begin{figure}
    \centering
    \includegraphics[width=1.0\textwidth]{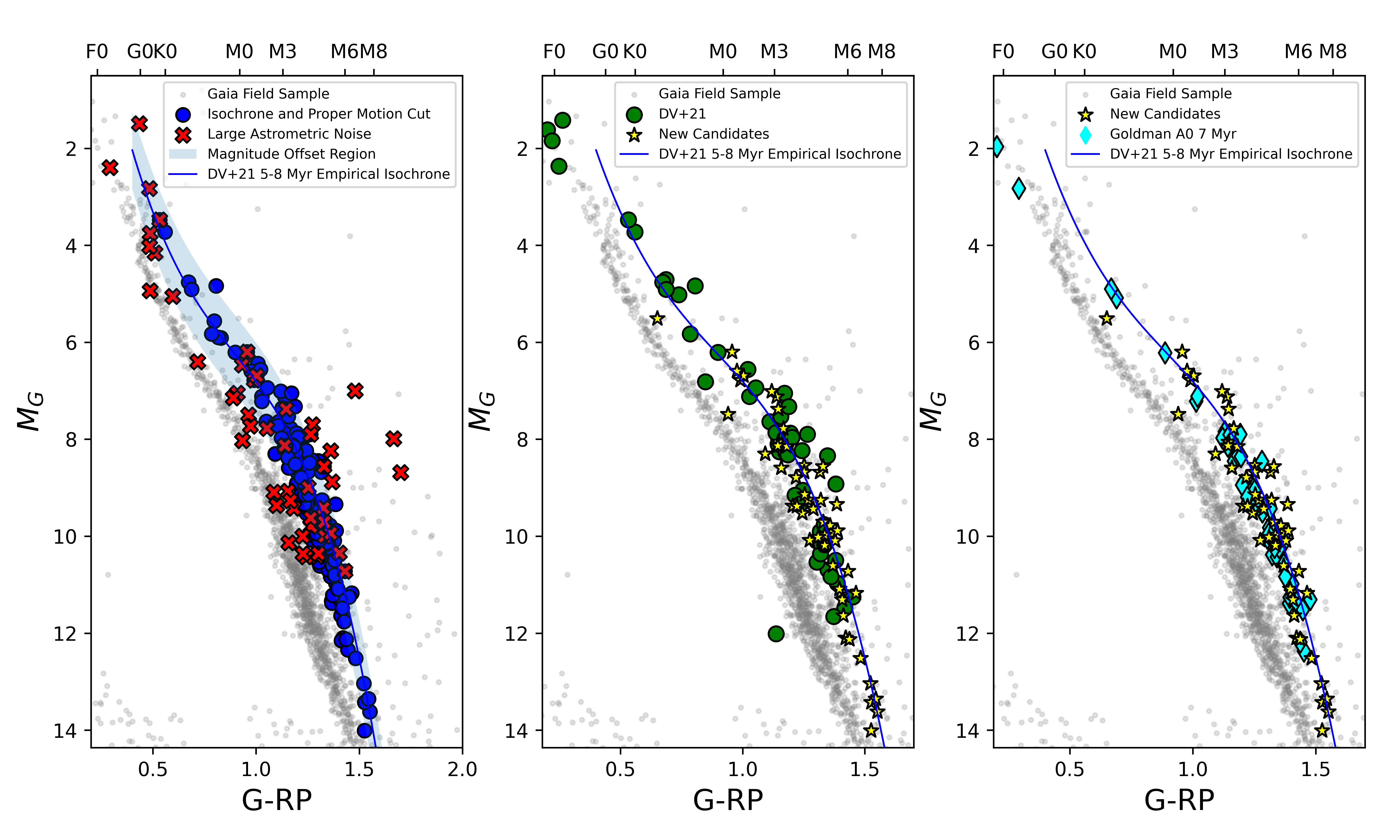}
    \caption{Left: Complete sample of $\sim$1600 stars satisfying initial RA and DEC search criteria (grey points). Blue points are stars that satisfy selection criteria for proper motion and distance from the empirical isochrone. Red X's are stars rejected on the basis of astrometric excess noise. Blue shading indicates the region used to select sample stars with ages close to the DV+21 5-8 Myr isochrone, shown as the blue curve. Middle: Comparison between the final sample meeting all candidate criteria (yellow star markers) and bona fide candidates from DV+21 in green. Right: Same final sample now compared with stars from the LCC subgroup A0 from \citep{goldman_large_2018} (teal diamonds).} 
    \label{fig:Three_CMD}
\end{figure}

To illustrate our successful retrieval of new ECA candidates, we present color-magnitude diagrams of the final sample in Fig.~\ref{fig:Three_CMD}. The left panel highlights the stars remaining after considering photometric quality and proper motion as well as their isochronal positions. The blue highlighted region shows the boundary defined by the magnitude offset in Eq.~\ref{eq:mag}; points marked with red X's show stars removed on the basis of astrometric quality. We are then left with the subset of stars expected to be young. 

In addition to identifying potential new ECA members, we also set out to independently evaluate the membership status of LCC stars with properties similar to those of ECA members. Thus, in the middle and right panels, we present (respectively) comparisons of our new ECA candidates with bona fide members of the ECA from DV+21 and members of the youngest LCC subgroup A0 from \citep[][hereafter G+18]{goldman_large_2018}. These comparisons demonstrate that our selected candidates are indeed young, i.e., of similar age to the LCC A0 subgroup and the ECA. Notably, there appear to be more previously identified ECA candidates located above the empirical isochrone, while the majority of our new candidate stars, as well as the A0 subgroup, sit below this curve. The latter stars are all late-type M dwarfs that appear to represent a population of age $\sim$8 Myr, i.e., near the upper bound of the empirical isochrone. 

While we refer to the Table \ref{table:gaia param} sample as consisting of our new ECA candidates (or, simply, ``our new candidates'') in the following discussion, we also continue to consider the possibility that these stars may belong to the younger LCC subgroups --- and, indeed, that there may be no clear delineation between ECA and young LCC members.
\begin{longrotatetable}
\begin{deluxetable*}{lCCCCCCCCCC}
\tablecaption{Gaia DR3 parameters and previous young star references for our New Candidates with Properties similar to the ECA\label{table:gaia param}}
\tablewidth{700pt}
\tabletypesize{\scriptsize}
\tablehead{
\colhead{Name} & \colhead{RA} & 
\colhead{DEC} & \colhead{$\pi$} & 
\colhead{$G$} & \colhead{$G-RP$} & 
\colhead{PMRA} & \colhead{PMDEC} & 
\colhead{RV} & \colhead{RUWE} & \colhead{Previous Ref.} \\ 
\colhead{} & \colhead{(deg)} & \colhead{(deg)} & \colhead{(mas)} & 
\colhead{mag} & \colhead{(mag)} & \colhead{(mas yr$^{-1}$)} &
\colhead{(mas yr$^{-1}$)} & \colhead{(km s$^{-1}$)} & \colhead{} & \colhead{}
} 
\startdata
Gaia DR3 5797741860004653696 & 214.6964 & -71.5458 & 9.6340  & 9.3743  & 1.2040 & -38.7725 & -9.4834  & ...               & 1.2044  & ...    \\
Gaia DR3 5797275387886033152 & 213.3555 & -73.4109 & 10.1316 & 9.9987  & 1.3330 & -43.2440 & -15.8041 & -1.81 $\pm$ 6.16  & 13.8351 & ...    \\
UCAC4 070-021428             & 211.1121 & -76.1450 & 9.5378  & 9.5234  & 1.2433 & -37.4574 & -15.5029 & 10.02 $\pm$ 4.15  & 1.1125  & 1      \\
Gaia DR3 5846840105074976384 & 209.0416 & -70.3259 & 10.7020 & 10.0829 & 1.2745 & -41.8814 & -18.9599 & ...               & 0.9824  & ...    \\
Gaia DR3 5841149170330789120 & 205.8611 & -69.7371 & 10.1434 & 8.3031  & 1.0921 & -42.8169 & -14.3138 & 11.23 $\pm$ 3.16  & 1.1291  & ...    \\
Gaia DR3 5791422313839597312 & 205.7002 & -73.9837 & 9.7629  & 9.9352  & 1.3612 & -39.1075 & -17.4423 & -1.88 $\pm$ 4.54  & 6.3264  & ...    \\
1RXS J132247.1-753231        & 200.6957 & -75.5432 & 10.4517 & 7.4852  & 0.9403 & -42.1503 & 0.6903   & 12.11 $\pm$ 9.46  & 0.9846  & 10     \\
JNN 92A                      & 199.9865 & -68.5206 & 10.3764 & 7.1212  & 1.1381 & -39.1164 & -18.4319 & ...               & 2.1369  & 1,2    \\
TYC 9250-1659-1              & 199.3370 & -72.9824 & 8.9833  & 5.5080  & 0.6503 & -37.9972 & 0.3588   & 13.14 $\pm$ 0.91  & 3.1982  & 10     \\
Gaia DR3 5840441634588412672 & 198.3425 & -72.0070 & 9.5670  & 12.0968 & 1.4213 & -39.3174 & -9.7597  & ...               & 1.0472  & ...    \\
UCAC4 105-061621             & 194.5726 & -69.1061 & 10.6449 & 10.1386 & 1.3320 & -41.9664 & -16.7836 & ...               & 1.0495  & 1,2,9  \\
2MASS J12560892-6926503      & 194.0368 & -69.4474 & 10.0289 & 11.1721 & 1.4635 & -37.2443 & -15.3647 & ...               & 1.3256  & 6      \\
1RXS J125608.8-692652        & 194.0341 & -69.4484 & 10.0585 & 6.1998  & 0.9563 & -41.7056 & -16.7780 & -15.35 $\pm$ 9    & 13.0557 & 3,6,10 \\
Gaia DR3 5856965846936162816 & 193.7715 & -68.7552 & 10.2987 & 10.0891 & 1.3783 & -41.7732 & -16.5954 & ...               & 14.4482 & ...    \\
UCAC4 068-016412             & 193.0619 & -76.4744 & 9.6342  & 9.2473  & 1.2675 & -37.5794 & -14.2828 & 14.68 $\pm$ 2.06  & 1.2016  & 1,10   \\
Gaia DR3 5837286379802373120 & 191.9880 & -76.1026 & 10.0044 & 10.1876 & 1.3356 & -40.2181 & -14.0581 & ...               & 1.1738  & 1      \\
Gaia DR3 5855802808484610304 & 191.6106 & -68.0866 & 9.7149  & 9.7764  & 1.3577 & -39.5959 & -13.3876 & ...               & 2.1028  & ...    \\
Gaia DR3 5843083383076334720 & 191.4015 & -70.8957 & 9.9059  & 8.6761  & 1.3161 & -38.4353 & -12.9981 & ...               & 2.6628  & ...    \\
ASAS J124413-6902.2          & 191.0603 & -69.0433 & 10.0437 & 7.3758  & 1.1453 & -38.2855 & -17.1387 & 9.28 $\pm$ 2.71   & 30.7021 & 3,10   \\
RX J1243.6-7834              & 190.9017 & -78.5689 & 9.9219  & 7.0115  & 1.1188 & -39.6004 & -13.2157 & 9.99 $\pm$ 3.6    & 7.3975  & 1,7,10 \\
UCAC4 105-056596             & 189.3904 & -69.0011 & 9.7765  & 8.5380  & 1.2516 & -38.4960 & -12.9136 & 16.14 $\pm$ 35.66 & 3.0364  & 1      \\
UCAC4 105-056549             & 189.3433 & -69.0685 & 9.3305  & 6.7814  & 0.9906 & -40.7237 & -10.6381 & 11.81 $\pm$ 0.87  & 11.0174 & 1      \\
Gaia DR3 5855362900730353792 & 189.2832 & -69.7701 & 9.5628  & 13.6179 & 1.5515 & -38.9198 & -12.2239 & ...               & 0.9708  & 5      \\
Gaia DR3 5788500121160692480 & 189.2594 & -78.1405 & 9.9639  & 14.0099 & 1.5258 & -38.3730 & -11.6673 & ...               & 1.0461  & 5      \\
1RXS J123652.9-760216        & 189.2442 & -76.0420 & 10.5428 & 9.8824  & 1.3875 & -41.8437 & 6.3454   & 21.37 $\pm$ 5.99  & 1.3117  & 10    \\
1RXS J123652.9-760216        & 189.2425 & -76.0424 & 10.6491 & 9.9758  & 1.3349 & -39.9747 & 9.4947   & 27.88 $\pm$ 4.72  & 1.2694  & 10    \\
UCAC4 081-025318             & 189.2070 & -73.9390 & 9.4922  & 7.7756  & 1.1663 & -41.2700 & -12.0132 & ...               & 3.0310  & 1,10   \\
UCAC4 109-056524             & 188.9877 & -68.3003 & 9.6017  & 9.7069  & 1.3296 & -38.0374 & -11.8874 & 14.15 $\pm$ 6.35  & 8.1737  & 2      \\
RX J1231.9-7848              & 187.9827 & -78.8091 & 10.0177 & 8.0571  & 1.1452 & -39.8105 & -11.5353 & ...               & 1.4916  & 1,7,10 \\
UCAC4 104-054759             & 187.4819 & -69.2693 & 9.5914  & 6.6886  & 1.0027 & -37.8974 & -9.8218  & 12.28 $\pm$ 3.74  & 10.6911 & 1      \\
Gaia DR3 5841358038860765568 & 187.3206 & -73.9116 & 9.9540  & 13.0368 & 1.5223 & -39.0290 & -10.6080 & ...               & 1.0560  & 5      \\
Gaia DR3 5842571697842877440 & 186.3942 & -71.9172 & 10.4799 & 11.1557 & 1.4016 & -41.2563 & -1.6421  & ...               & 1.1503  & 1      \\
Gaia DR3 5841625739885718144 & 185.8739 & -73.1740 & 9.6570  & 11.6371 & 1.4116 & -39.6831 & -9.7932  & ...               & 2.9365  & ...    \\
Gaia DR3 5856254531639542144 & 184.1161 & -68.9656 & 9.5038  & 13.4261 & 1.5265 & -39.2797 & -9.5702  & ...               & 1.0034  & ...    \\
Gaia DR3 5837841117779242496 & 183.8567 & -76.0584 & 9.4131  & 10.6050 & 1.3687 & -38.2017 & -7.8234  & 26.66 $\pm$ 4.32  & 1.4578  & ...    \\
ASAS J121413-7321.6          & 183.5717 & -73.3595 & 9.8603  & 6.5950  & 0.9762 & -39.3654 & -11.0226 & ...               & 8.2704  & 1,10   \\
UCAC4 054-011484             & 183.5189 & -79.2311 & 9.9830  & 9.4348  & 1.2888 & -41.6995 & -7.7002  & 15.86 $\pm$ 2.37  & 1.1875  & 1      \\
UCAC4 059-012851             & 183.0943 & -78.3678 & 9.8383  & 9.3394  & 1.3848 & -40.6920 & -7.2419  & ...               & 1.3888  & 1      \\
eps cha 12                   & 181.9407 & -78.2685 & 9.3698  & 9.3803  & 1.2215 & -38.6229 & -6.4996  & 11.84 $\pm$ 2.86  & 1.1127  & 11     \\
Gaia DR3 5837882207728535040 & 181.1041 & -76.0451 & 9.6117  & 13.3564 & 1.5443 & -40.9589 & -6.4138  & ...               & 1.0990  & ...    \\
UCAC4 105-048794             & 180.7584 & -69.0869 & 10.3191 & 8.5897  & 1.1580 & -38.4681 & -0.6930  & ...               & 1.1553  & 2      \\
2MASS J12003792-7845082      & 180.1571 & -78.7523 & 9.8383  & 11.3104 & 1.4067 & -41.6640 & -6.0499  & ...               & 1.1071  & 8      \\
UCAC4 056-012157             & 180.0037 & -78.8081 & 9.8176  & 8.6757  & 1.2509 & -41.3636 & -5.4022  & ...               & 1.2750  & 1      \\
Gaia DR3 5227167240115285248 & 179.8743 & -72.6405 & 9.8538  & 10.0210 & 1.3089 & -40.7344 & -5.5530  & 11.02 $\pm$ 4.55  & 1.1536  & 1      \\
Gaia DR3 5200024318108381568 & 178.3925 & -79.0425 & 9.7131  & 8.5585  & 1.3289 & -40.6353 & -5.6043  & 24.49 $\pm$ 18.18 & 20.8155 & ...    \\
UCAC4 096-034374             & 178.1574 & -70.9835 & 9.7386  & 8.1459  & 1.1811 & -37.2360 & -2.8336  & 12.14 $\pm$ 22.83 & 1.3589  & 1      \\
Gaia DR3 5233289252143166336 & 178.0155 & -71.5468 & 9.5740  & 12.5182 & 1.4824 & -39.8541 & -3.4165  & ...               & 0.9934  & ...    \\
Gaia DR3 5227282517035989632 & 175.7426 & -72.8147 & 10.1551 & 10.7195 & 1.4307 & -37.5608 & -3.6890  & 15.14 $\pm$ 4.37  & 4.4240  & ...    \\
UCAC4 060-011194             & 173.5396 & -78.0015 & 9.9702  & 8.1344  & 1.1436 & -42.1001 & -0.9880  & 14.32 $\pm$ 4.31  & 1.1785  & 1,10   \\
UCAC4 086-024494             & 167.7166 & -72.9203 & 9.8783  & 9.2522  & 1.3196 & -38.9016 & 1.0986   & 20.31 $\pm$ 4.57  & 1.6128  & 1      \\
UCAC4 101-037476             & 166.0989 & -69.8227 & 10.0553 & 8.7858  & 1.2176 & -40.6668 & 2.5467   & ...               & 1.4490  & 1,2    \\
2MASS J10563146-7618334      & 164.1304 & -76.3093 & 10.0337 & 9.1373  & 1.2533 & -40.0500 & 5.2798   & 15.17 $\pm$ 3.42  & 1.1711  & 1,12   \\
Gaia DR3 5199332484776169600 & 159.4241 & -78.8193 & 10.4768 & 11.0867 & 1.3954 & -41.9504 & 9.9341   & ...               & 1.0482  & 1      \\
Gaia DR3 5229663337667889152 & 158.4319 & -71.7612 & 9.9938  & 12.1279 & 1.4361 & -39.6886 & 8.2466   & ... & 1.0169  & ...
\enddata
\tablecomments{Previous references identifying either group membership or signs of youth:(1) \citet{zari_3d_2018}, (2) \citet{gagne_banyan_2018}, (3) \citet{song_new_2012}, (4) \citet{kiraga_asas_2012}, (5) \citet{reyle_new_2018}, (6) \citet{bohn_unveiling_2022}, (7) \citet{alcala_study_1997}, (8) \citet{schutte_discovery_2020}, (9) \citet{goldman_large_2018}, (10) \citet{voges_vizier_2000}, (11) \citet{dickson-vandervelde_gaia-based_2021}, (12) \citet{frasca_gaia-eso_2015}}
\end{deluxetable*}
\end{longrotatetable}

\subsection{Candidate Spatio-Kinematics}\label{sec:Kinematic}

To establish the relationships between the spatial distributions and kinematics of our new ECA candidates vs. those of previously identified ECA and LCC (A0) members and candidates, we determine each star's Galactic-heliocentric position ($XYZ$) and velocity ($UVW$).  Galactic-heliocentric coordinates are defined as the Sun's galactic position representing the origin, with the $X$ axis pointing towards the galactic center, $Y$ being perpendicular to $X$ having a positive direction towards galactic rotation, and $Z$ perpendicular to both $X$ and $Y$ with positive direction being the same as the galactic north pole. Each axis has a corresponding velocity component of $UVW$. Due to a lack of radial velocity measurements, especially for the faintest candidate stars, not every star has a complete set of $UVW$ kinematics. Specifically, only 26 out of 54 stars have radial velocities from Gaia DR3. Analysis of $UVW$ velocities will therefore only pertain to stars that have measured radial velocities. To obtain a more complete kinematic view of our candidates, we also use tangential velocities in RA and Declination as derived from proper motion measurements, where these tangential velocity components are defined (respectively) as $V_{\perp,\alpha} = \kappa \mu_{\alpha}/\varpi$, $V_{\perp,\delta} =\kappa \mu_{\delta}/\varpi$ with $\kappa =  4.74047$. 

\begin{figure}
    \centering
    \includegraphics[width = 0.8\textwidth]{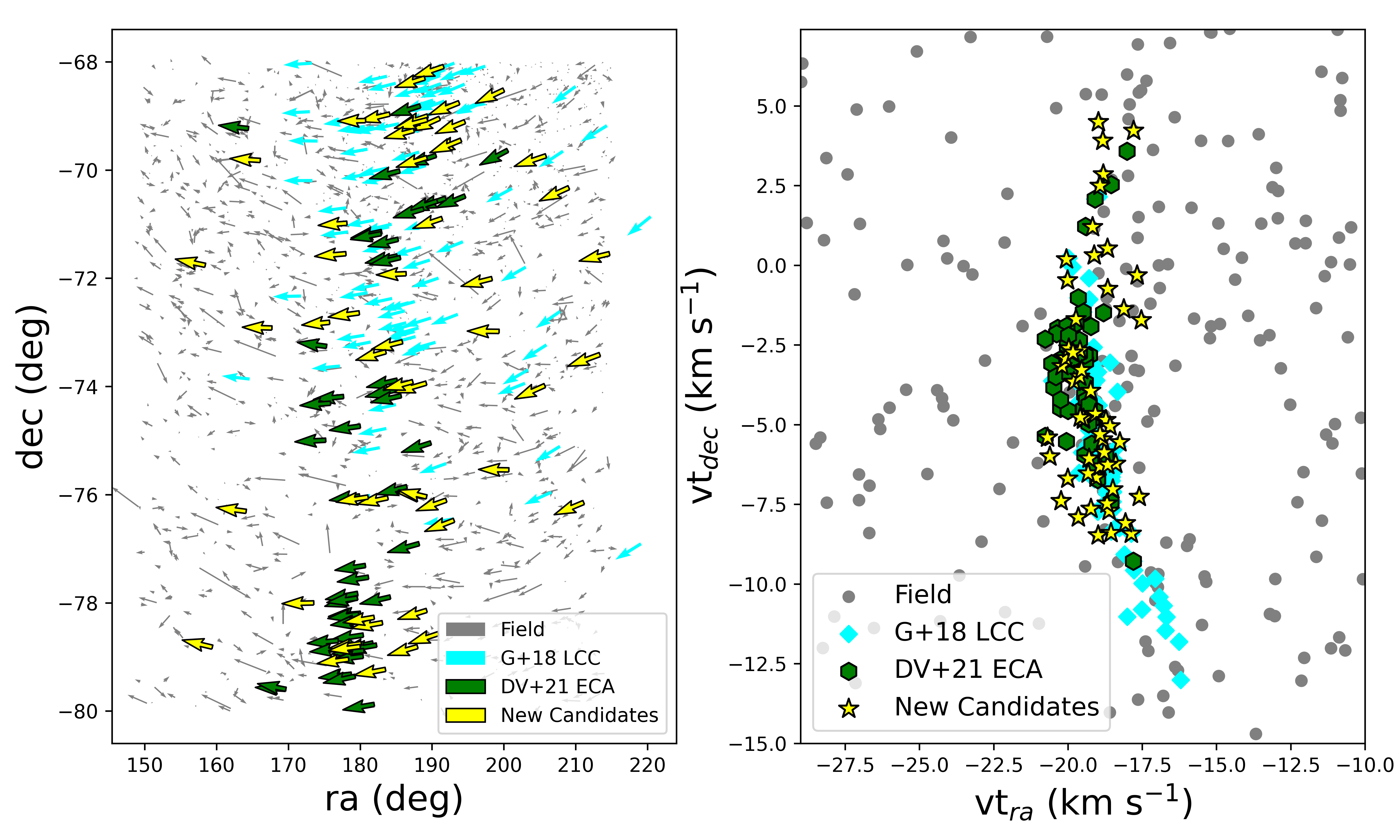}
    \caption{Left: The sky position and tangential velocity vectors of the new ECA candidates (\S \ref{sec:selection} and Table \ref{table:gaia param}), LCC A0 subgroup, and previous ECA members. Right: The tangential velocity vectors of the the new ECA candidates, LCC A0 subgroup, and previous ECA members.}
    \label{fig:tangv}
\end{figure}

In Fig.~\ref{fig:tangv}, we display the spatial coordinates and tangential velocity vectors of our final sample of new ECA candidates, along with those of known ECA and LCC members. It is apparent that the final sample has sky positions and space motions similar to both the ECA and LCC, and that the velocity vectors of the candidates and ECA and LCC members are distinct from those of the field stars in their vicinity. This map also highlights the near-complete ($G \lesssim 15$ mag) Gaia DR3 sample of co-moving stars in this $50 \times 50 \times 50$ pc$^3$ cube of space located $\sim$100 pc from the Sun. As an overview of the positions and the motions of our new candidates, we list the average heliocentric spatio-kinematics coordinates for our new candidates, DV+21 ECA members, and G+18 LCC A0 members in Table \ref{table:motion}.

\begin{deluxetable*}{lCCCCCCCC}
\tablecaption{\sc New and previously identified ECA and LCC candidates: mean positions and velocities \label{table:motion}}
\tablewidth{0pt}
\tablehead{
\colhead{Group} & \colhead{$X$} &  \colhead{$Y$} & \colhead{$Z$} & \colhead{$U$} & \colhead{$V$} & \colhead{$W$} \\
\colhead{} & \colhead{(pc)} & \colhead{(pc)} & \colhead{(pc)} & \colhead{(km s$^{-1}$)} & \colhead{(km s$^{-1}$)} & \colhead{(km s$^{-1}$)}
}

\startdata
New ECA Candidates & 52.248 & -86.935 & -16.661 & -10.493 & -19.306 & -9.817\\
Established ECA Candidates (DV+21) & 49.669	 & -85.176 & -23.745 & -9.847 & -20.667 & -9.682\\
LCC A0 (G+18) & 52.187 & -87.977 & -17.222 & -10.352 & -19.018 & -8.776\\
\enddata

\end{deluxetable*}

\section{Results: Candidate Properties}\label{sec:section 3}

\subsection{Spatial and kinematic distributions}\label{sec:kinematics}
\begin{figure}
    \centering
    \includegraphics[width = 0.8\textwidth]{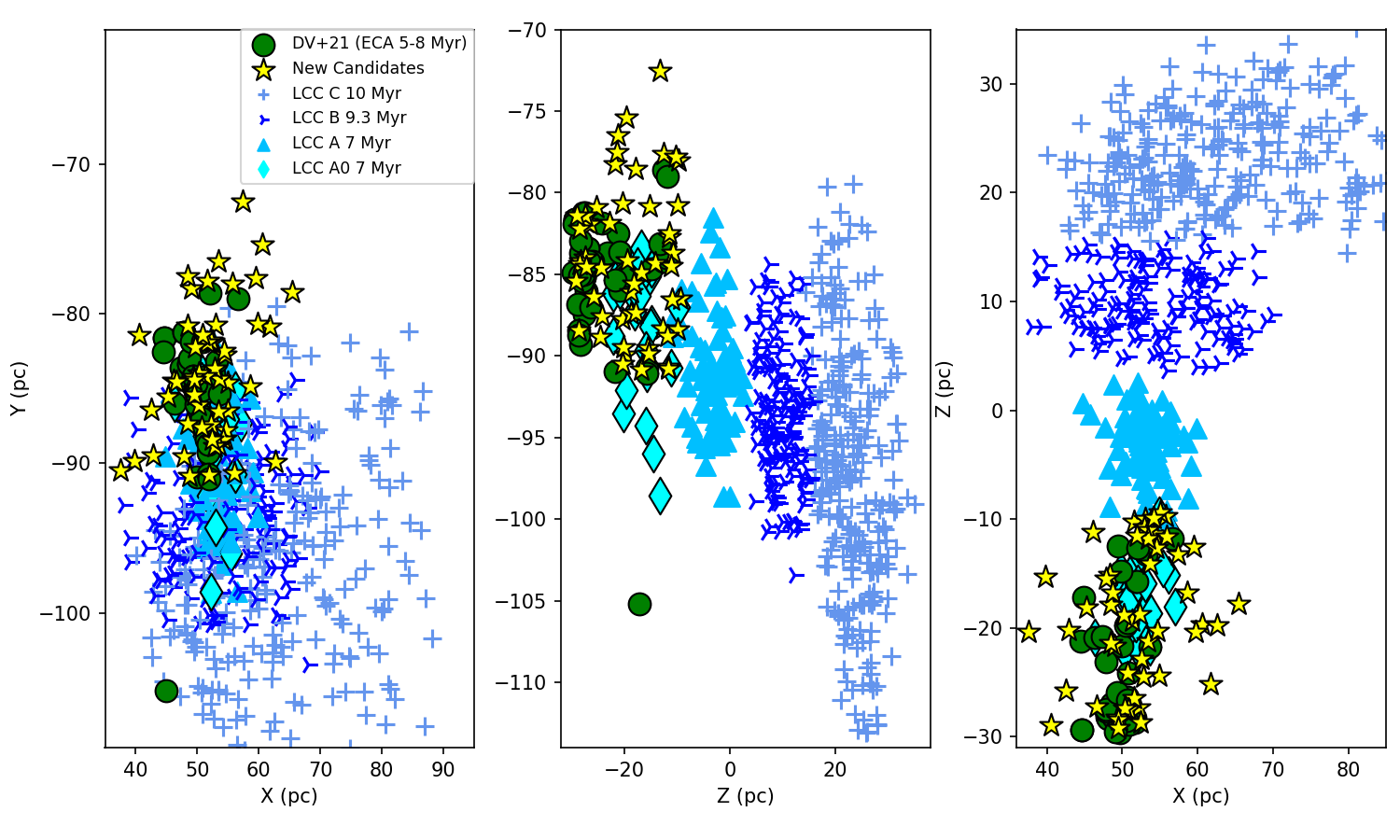}
    \caption{Heliocentric positions of our sample as yellow stars, the LCC subgroups from G+18 in different shades of blue, and bona fide ECA members in green. Our sample shows new stars scattered around the ECA core as well as around the A0 subgroup. We also recovered existing ECA and A0 stars.}
    \label{fig:position_comparison}
\end{figure}

\begin{figure}
    \centering
    \includegraphics[width = 0.8\textwidth]{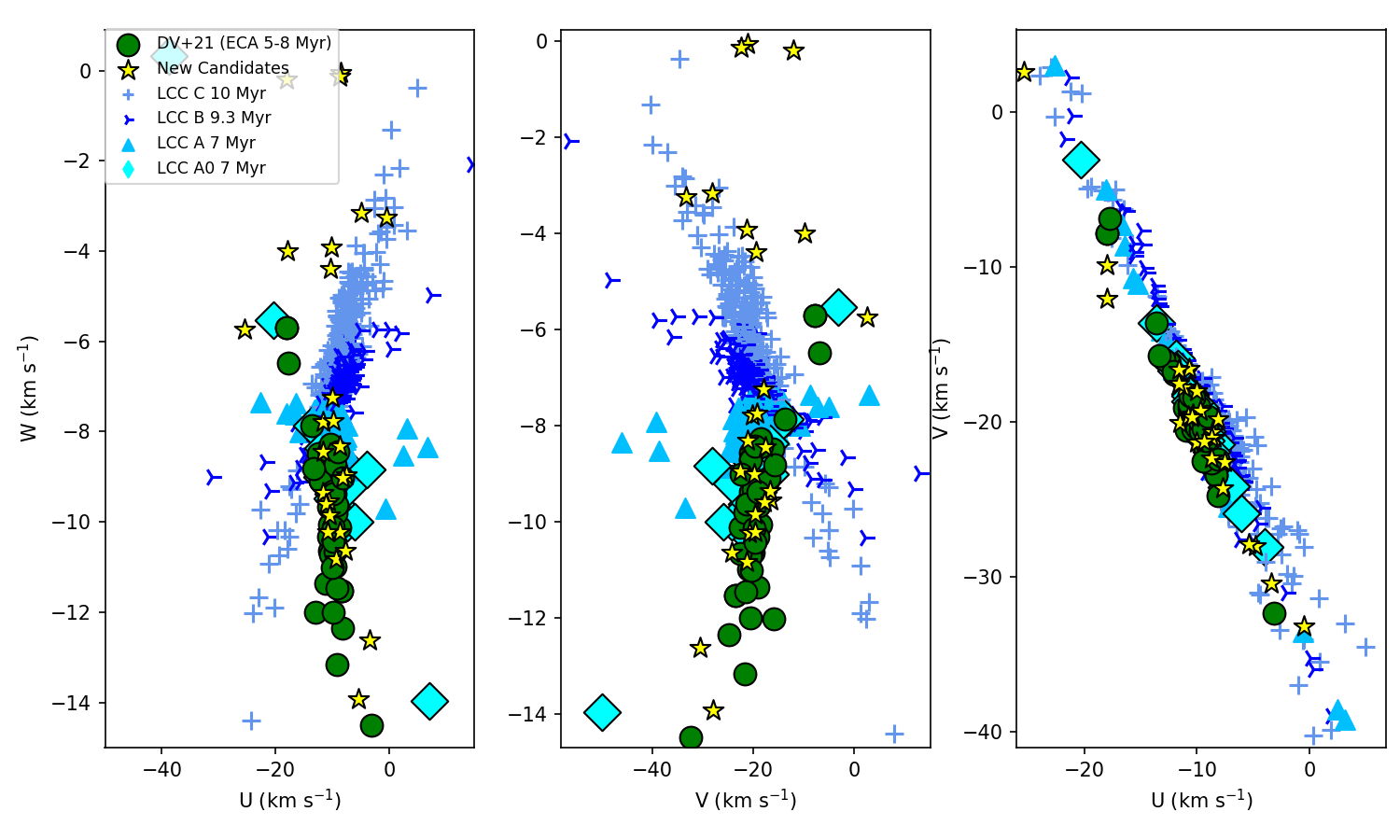}
    \caption{Heliocentric velocities shown in all three directions. The LCC subgroups from G+18 in different shades of blue, and bona fide ECA members (DV+21) in green. New candidates stars are in yellow. LCC A0 subgroup stars are in blue diamonds.}
    \label{fig:velocity_comparison}
\end{figure}

In Fig.~\ref{fig:position_comparison} and Fig.~\ref{fig:velocity_comparison}, we present spatial and kinematic maps of our candidates, along with LCC stars from G+18 and members of the ECA as established by DV+21. Fig. \ref{fig:position_comparison} shows each spatial plane in $XYZ$ coordinates for these stars. The LCC (blue symbols) is seen as a large association spanning $\sim$60 pc. As described in G+18, there exist several LCC subgroups that correlate with age, in the form of a sequence from $\sim$20 Myr to $\sim$8 Myr. At the `south' end (negative $Z$ positions), the youngest stars in the LCC form the A0 subgroup (blue diamond markers). However ECA stars (green circles) start to occupy the same region. These ECA stars have an upper bound age of $\sim$8 Myr, very similar to the average age for the A0 stars. 
Our selected candidates show positions consistent both with older ECA stars as well as LCC A0 stars. We also see that several of our new candidates are found in the ECA `core' ($XYZ$ coordinates 50, $-$85, $-$29). 

Fig. ~\ref{fig:velocity_comparison} shows the $UVW$ motion of each group. At first glance, older LCC subgroup members show distinct clustering, appearing aligned in a positive $UW$ trend and negative $VW$ trend. Meanwhile, the younger A and A0 subgroups of the LCC, along with the ECA members, display the opposite, with a different cluster of motions. In terms of kinematics ($UVW$), our new candidates clearly best match the ECA and youngest (A0) LCC subgroup. Furthermore, the LCC A0 subgroup, the ECA, and our new candidates display very similar average velocities (Table \ref{table:motion}). 

\subsection{SED Fitting: Candidate Stellar Parameters} \label{subsec3.2}

By combining Gaia DR3 $G_{rp}$, $G_{bp}$, and $G$ photometry with data from the 2MASS \citep{skrutskie_two_2006} and WISE \citep{wright_wide-field_2010} infrared catalogues, we can compile spectral energy distributions (SEDs) that cover most of a cool star's photospheric emission. With the numerous and large libraries of synthetic model atmospheres, it is possible to fit synthetic photometry derived from these models to the resulting photospheric SEDs. 
Using the VOSA \citep{bayo_vosa_2008} web 
application\footnote{\url{http://svo2.cab.inta-csic.es/theory/vosa/help/star/intro/}}, we extracted 
calibrated photometry from the Gaia DR3, 2MASS, and WISE catalogues to generate observed SEDs for each star. We then used the VOSA web tool to fit stellar atmosphere models to each SED and thereby to determine each star's bolometric luminosity and obtain estimates for stellar radius, mass, temperature, and age. 
The VOSA tool fits synthetic model photometry for the Gaia/2MASS/WISE passbands to the observed SEDs, with the best fit determined by minimizing the reduced chi-squared value across a grid of stellar atmosphere model fits. Mid-IR (WISE) photometry that significantly deviates from the synthetic model photometry is ignored during the fitting process.

\begin{figure}
    \centering
    \includegraphics[width = 1.0\textwidth]{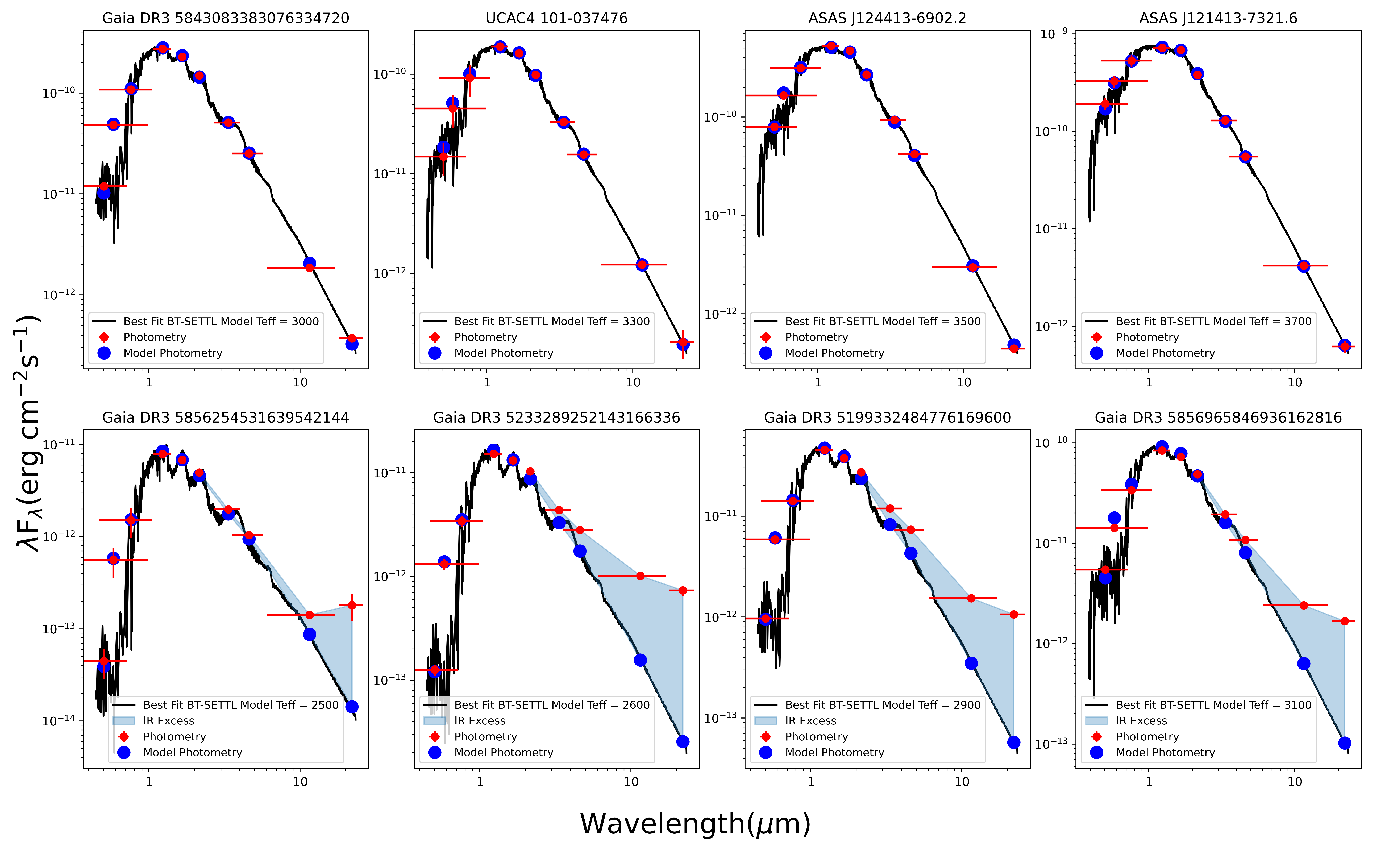}
    \caption{Eight sample VOSA SED fits from among the fits for our 54 new candidates. Observed photometry from Gaia DR3, 2MASS, and WISE is presented as blue points, where the horizontal bars indicate the passbands. The best-fit BT-SETTL models 
    are shown in black, with 
    red points indicating the synthetic model photometry used in the fitting process. The top row shows four stars with no measurable infrared excess; the bottom row shows a sample of four stars with significant infrared excess, as indicated by the light blue highlighted regions. 
    }
    \label{fig:sed fit}
\end{figure}

The stellar parameters (and their formal uncertainties\footnote{The (uniform) formal uncertainties in effective temperatures obtained from the VOSA fitting, $\pm$50 K, are essentially determined by the stellar model grid spacing.}) obtained from the VOSA SED fits are listed in Table~\ref{table:estimated params}. Representative examples of the resulting VOSA fits of BT-SETTL model atmospheres to Gaia/2MASS/WISE SEDs of our new candidates are presented in Fig.~\ref{fig:sed fit}.  The bottom panels of Fig.~\ref{fig:sed fit} show stars with evidence of infrared excess, based on deviations from the model-predicted (synthetic) 2MASS and WISE photometry (see Sec.~\ref{sec:IRexcesses}).  

To derive stellar properties for all of the stars in our sample, as well as for the stars found in G+18's LCC A0 subgroup and DV+21's ECA members, we chose two classes of models: the BT-SETTLE CFIST  models \citep{baraffe_new_2015} with solar metallicities and, for purposes of better constraining stellar age estimates, the SPOTS models \citep{somers_spots_2020}. The latter model class accounts for the potential effects of magnetic activity in late-type stars, such as heavy stellar surface spotting and an inflated stellar atmosphere. Comparison of results from these two model classes is potentially informative, given that the majority of new candidate young stars found in the lower LCC and ECA region have been low-mass, late-type (K and M) dwarfs \citep{goldman_large_2018,dickson-vandervelde_gaia-based_2021,kubiak_new_2021}; indeed, recent studies have established that one must account for the effects of a magnetized atmosphere so as to avoid underestimating ages when modeling the SEDs of such young, magnetically active late-K and M dwarfs \citep{somers_older_2015,somers_spots_2020}. To enable comparisons between the stellar properties we derive for our new ECA candidates and those of previously identified LCC stars, we used the aforementioned models to estimate the effective temperatures, luminosities, and ages of all LCC member stars from \citet{goldman_large_2018} that have declinations $\delta < -68$. This declination limit includes all previously identified stars in the ECA and the LCC A0 subgroup, as well as portions of the A and Z subgroups in the LCC.
 
\begin{figure}
    \centering
    \includegraphics[width = 1.0\textwidth]{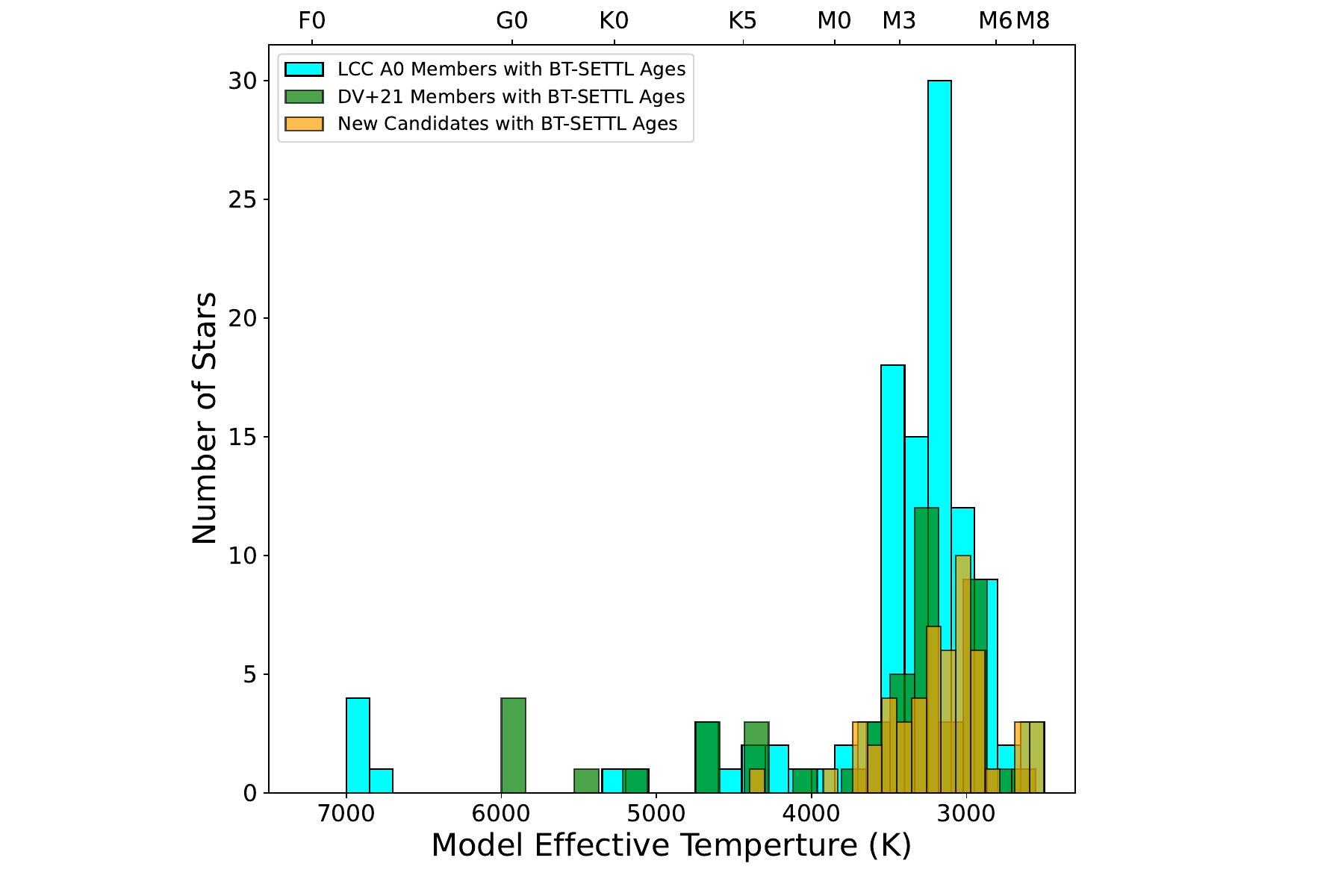}
    \caption{BT-SETTLE best-fit temperatures and derived spectral types \citep{pecaut_intrinsic_2013} for new sample candidates (yellow), LCC A0 members (teal) and ECA members (green).}
    \label{fig:spt hist}
\end{figure}

Fig. \ref{fig:spt hist} displays a histogram of effective temperatures (from best fit BT-Settle models) and the corresponding spectral types \citep[as obtained from Table 5][]{pecaut_intrinsic_2013} for our new ECA candidates. The figure demonstrates that the majority of stars in our new ECA candidate sample are M dwarfs. The peak in the spectral type distribution of our candidates (around M5) is somewhat later than the peaks in the spectral type distributions found for ECA stars in DV+21 and LCC stars in G+18 (both around M3.5). 
Thus, with Gaia DR3, we are able to add several dozen late-type M dwarfs and even young proto-brown dwarf candidates to the ECA/LCC spatio-kinematic overlap region. 

\begin{longrotatetable}
\begin{deluxetable*}{lCCCCCCCCCCC}
\tablecaption{New Candidates derived kinematics, positions, and estimated paramaters\label{table:estimated params}}
\tablewidth{700pt}
\tabletypesize{\scriptsize}
\tablehead{
\colhead{Name} & \colhead{Distance} & 
\colhead{$U$} & \colhead{$V$} & 
\colhead{$W$} & \colhead{$X$} & 
\colhead{$Y$} & \colhead{$Z$} & 
\colhead{LogL} & \colhead{$T_{eff}$} & \colhead{Age (BT-Settl)} & \colhead{Age (SPOTS)} \\ 
\colhead{} & \colhead{(pc)} & \colhead{(pc)} & \colhead{(pc)} & 
\colhead{(pc)} & \colhead{(km s$^{-1}$)} & \colhead{(km s$^{-1}$)} &
\colhead{(km s$^{-1}$)} & \colhead{(L$_{\odot}$)} & \colhead{(K)} & \colhead{(Myr)} & \colhead{(Myr)}
} 
\startdata
Gaia DR3 5797741860004653696 & 103.796 & ...                & ...                & ...               & 65.46 & -78.58 & -17.75 & $-1.45^{+0.002}_{0.002}$  & 3300   & $12.94^{+6.72}_{-3.66}$ & $34.95^{+9.62}_{-8.08}$  \\
Gaia DR3 5797275387886033152 & 98.698  & -17.98 $\pm$ 3.78  & -12.03 $\pm$ 4.71  & -0.21 $\pm$ 1.23  & 60.59 & -75.4  & -19.64 & $-1.666^{+0.002}_{0.002}$ & 3300   & $21.7^{+8.3}_{-3.22}$   & ...                      \\
UCAC4 070-021428             & 104.843 & -10.34 $\pm$ 2.45  & -19.51 $\pm$ 3.2   & -4.38 $\pm$ 1     & 61.76 & -80.89 & -25.2  & $-1.45^{+0.003}_{0.004}$  & 3200   & $6.83^{+2.87}_{-1.83}$  & $10.48^{+3.62}_{-2.45}$  \\
Gaia DR3 5846840105074976384 & 93.437  & ...                & ...                & ...               & 57.42 & -72.52 & -13.23 & $-1.652^{+0.005}_{0.004}$ & 3200   & $10.46^{+7.6}_{-2.43}$  & $22.45^{+10.25}_{-6.57}$ \\
Gaia DR3 5841149170330789120 & 98.583  & -10.07 $\pm$ 1.9   & -21.33 $\pm$ 2.49  & -3.92 $\pm$ 0.4   & 59.44 & -77.64 & -12.56 & $-1.119^{+0.003}_{0.004}$ & 3600   & $26.99^{+4.72}_{-6.99}$ & $1.8^{+0.26}_{-0.22}$    \\
Gaia DR3 5791422313839597312 & 102.426 & -17.95 $\pm$ 2.65  & -9.9 $\pm$ 3.58    & -4 $\pm$ 0.91     & 59.74 & -80.67 & -20.37 & $-1.521^{+0.002}_{0.002}$ & 3000   & $4.17^{+0.04}_{-0.2}$   & $15.83^{+5.38}_{-4.05}$  \\
1RXS J132247.1-753231        & 95.674  & -8.42 $\pm$ 5.28   & -21.02 $\pm$ 7.57  & -0.06 $\pm$ 2.1   & 53.42 & -76.49 & -21.19 & $-0.885^{+0.04}_{0.043}$  & 3900   & $37.61^{+0.51}$         & $12.86^{+4.64}_{-2.86}$  \\
** JNN 92A                   & 96.369  & ...                & ...                & ...               & 55.74 & -78.01 & -9.73  & $-0.618^{+0.009}_{0.01}$  & 3500   & $3.1^{+0.97}_{-0.48}$   & $11.05^{+3.15}_{-2.63}$  \\
TYC 9250-1659-1              & 111.314 & -8.74 $\pm$ 0.51   & -22.34 $\pm$ 0.74  & -0.13 $\pm$ 0.16  & 62.61 & -89.9  & -19.73 & $-0.317^{+0.024}_{0.025}$ & 4400   & $19.99^{+2.71}_{-1.44}$ & $6.9^{+1.96}_{-2.45}$    \\
Gaia DR3 5840441634588412672 & 104.523 & ...                & ...                & ...               & 58.66 & -84.88 & -16.73 & $-2.258^{+0.018}_{0.017}$ & 2800   & $9.56^{+3.74}_{-2.57}$  & $1.26^{+0.55}$           \\
UCAC4 105-061621             & 93.938  & ...                & ...                & ...               & 51.6  & -77.83 & -10.22 & $-1.602^{+0.002}_{0.002}$ & 3000   & $4.85^{+0.29}_{-0.87}$  & $4.98^{+0.77}_{-0.92}$   \\
2MASS J12560892-6926503      & 99.709  & ...                & ...                & ...               & 54.45 & -82.74 & -11.43 & $-1.879^{+0.014}_{0.015}$ & 3000   & $8.87^{+2}_{-2.65}$     & ...                      \\
1RXS J125608.8-692652        & 99.415  & -25.41 $\pm$ 4.92  & 2.57 $\pm$ 7.47    & -5.74 $\pm$ 1.04  & 54.29 & -82.5  & -11.39 & $-0.338^{+0.007}_{0.006}$ & 3600   & $2.1^{+0.85}_{-0.14}$   & $2.99^{+0.56}_{-0.47}$   \\
Gaia DR3 5856965846936162816 & 97.096  & ...                & ...                & ...               & 52.98 & -80.76 & -9.96  & $-1.548^{+0.002}_{0.002}$ & 3100   & $4.97^{+1.18}_{-0.48}$  & $5.62^{+1.46}_{-1.15}$   \\
UCAC4 068-016412             & 103.794 & -8.66 $\pm$ 1.09   & -20.68 $\pm$ 1.68  & -10.23 $\pm$ 0.49 & 54.92 & -84.63 & -24.41 & $-1.319^{+0.003}_{0.004}$ & 3200   & $5^{+1.93}_{-0.99}$     & $5.65^{+1.42}_{-0.77}$   \\
Gaia DR3 5837286379802373120 & 99.953  & ...                & ...                & ...               & 52.59 & -81.86 & -22.88 & $-1.614^{+0.002}_{0.002}$ & 3000   & $5.22^{+0.03}_{-1.22}$  & ...                      \\
Gaia DR3 5855802808484610304 & 102.931 & ...                & ...                & ...               & 55.02 & -86.49 & -9.36  & $-1.426^{+0.01}_{0.01}$   & 3000   & $2.93^{+0.22}$          & ...                      \\
Gaia DR3 5843083383076334720 & 100.947 & ...                & ...                & ...               & 53.63 & -84.35 & -14.1  & $-1.027^{+0.002}_{0.003}$ & 3000   & ...                     & $11.2^{+3.23}_{-2.48}$   \\
ASAS J124413-6902.2          & 99.562  & -10.66 $\pm$ 1.45  & -16.55 $\pm$ 2.28  & -9.55 $\pm$ 0.37  & 52.87 & -83.68 & -10.72 & $-0.703^{+0.002}_{0.003}$ & 3500   & $4.07^{+1.15}_{-0.96}$  & $20.32^{+7.81}_{-4.7}$   \\
RX J1243.6-7834              & 100.784 & -11.62 $\pm$ 1.87  & -16.6 $\pm$ 2.93   & -9.36 $\pm$ 0.98  & 52.17 & -81.8  & -27.28 & $-0.564^{+0.009}_{0.008}$ & 3500   & $2.92^{+0.71}_{-0.86}$  & $5.12^{+1.25}_{-0.85}$   \\
UCAC4 105-056596             & 102.283 & -7.57 $\pm$ 18.62  & -22.61 $\pm$ 30.17 & -8.96 $\pm$ 3.83  & 53.41 & -86.54 & -10.98 & $-1.055^{+0.009}_{0.01}$  & 3200   & $2.99^{+0.88}_{-0.86}$  & $12.58^{+5.09}_{-3.46}$  \\
UCAC4 105-056549             & 107.172 & -11.55 $\pm$ 0.46  & -20.08 $\pm$ 0.74  & -7.78 $\pm$ 0.13  & 55.93 & -90.68 & -11.63 & $-0.573^{+0.002}_{0.003}$ & 3700   & $6.88^{+1.12}_{-1.19}$  & $12.54^{+4}_{-3.36}$     \\
Gaia DR3 5855362900730353792 & 104.569 & ...                & ...                & ...               & 54.53 & -88.33 & -12.62 & $-2.592^{+0.003}_{0.003}$ & 2500   & $8.7^{+1.67}_{-5.79}$   & $20.04^{+7.77}_{-4.75}$  \\
Gaia DR3 5788500121160692480 & 100.359 & ...                & ...                & ...               & 51.54 & -81.95 & -26.47 & $-2.721^{+0.004}_{0.003}$ & 2600   & $19.65^{+1.12}_{-2.79}$ & $4.47^{+1.4}_{-0.92}$    \\
1RXS J123652.9-760216        & 94.848  & -4.82 $\pm$ 3.1    & -28.03 $\pm$ 4.95  & -3.16 $\pm$ 1.37  & 48.99 & -78.28 & -21.65 & $-1.433^{+0.003}_{0.004}$ & 3100   & $4^{+0.99}_{-0.93}$     & ...                      \\
1RXS J123652.9-760216        & 93.901  & -0.49 $\pm$ 2.44   & -33.18 $\pm$ 3.9   & -3.25 $\pm$ 1.08  & 48.5  & -77.5  & -21.44 & $-1.455^{+0.004}_{0.003}$ & 3200   & $6.92^{+3.05}_{-1.92}$  & $14.19^{+5.62}$          \\
UCAC4 081-025318             & 105.347 & ...                & ...                & ...               & 54.63 & -87.76 & -20.28 & $-0.815^{+0.012}_{0.012}$ & 3400   & $3.62^{+1.21}_{-0.69}$  & $49.06^{+0.51}$          \\
UCAC4 109-056524             & 104.145 & -8.68 $\pm$ 3.3    & -21.05 $\pm$ 5.39  & -8.31 $\pm$ 0.61  & 54.16 & -88.4  & -9.93  & $-1.474^{+0.016}_{0.016}$ & 3100   & $4.16^{+1.35}_{-0.62}$  & ...                      \\
RX J1231.9-7848              & 99.82   & ...                & ...                & ...               & 50.78 & -81.43 & -27.47 & $-0.944^{+0.022}_{0.023}$ & 3500   & $9^{+2.26}_{-2.77}$     & $16.21^{+6.2}_{-3.69}$   \\
UCAC4 104-054759             & 104.257 & -9.74 $\pm$ 1.91   & -19.25 $\pm$ 3.18  & -7.76 $\pm$ 0.43  & 53.38 & -88.78 & -11.76 & $-0.514^{+0.021}_{0.023}$ & 3700   & $5.82^{+1.29}_{-1.18}$  & $4.14^{+1}_{-0.65}$      \\
Gaia DR3 5841358038860765568 & 100.459 & ...                & ...                & ...               & 51.32 & -84.17 & -19.35 & $-2.421^{+0.003}_{0.002}$ & 2500   & $1.99^{+3.02}_{-0.77}$  & $9.45^{+3.03}_{-2.02}$   \\
Gaia DR3 5842571697842877440 & 95.417  & ...                & ...                & ...               & 48.38 & -80.83 & -15.17 & $-1.931^{+0.002}_{0.003}$ & 2900   & $5.87^{+2.12}$          & $1.26^{+0.67}$           \\
Gaia DR3 5841625739885718144 & 103.549 & ...                & ...                & ...               & 52.27 & -87.41 & -18.72 & $-1.905^{+0.003}_{0.003}$ & 3100   & $15.11^{+4.07}_{-4.06}$ & $18.54^{+6.57}_{-4.42}$  \\
Gaia DR3 5856254531639542144 & 105.218 & ...                & ...                & ...               & 51.95 & -90.77 & -11.56 & $-2.569^{+0.003}_{0.003}$ & 2500   & $7.14^{+2.96}_{-4.72}$  & $39.94^{+0.52}$          \\
Gaia DR3 5837841117779242496 & 106.232 & -3.43 $\pm$ 2.15   & -30.42 $\pm$ 3.61  & -12.63 $\pm$ 1    & 52.79 & -88.87 & -24.52 & $-1.767^{+0.002}_{0.002}$ & 3000   & $6.36^{+1.66}_{-1.34}$  & ...                      \\
ASAS J121413-7321.6          & 101.413 & ...                & ...                & ...               & 50.18 & -86.1  & -18.8  & $-0.496^{+0.012}_{0.012}$ & 3700   & $5.32^{+1.46}_{-1.22}$  & $20.03^{+5.24}_{-4.12}$  \\
UCAC4 054-011484             & 100.167 & -9.41 $\pm$ 1.17   & -21.25 $\pm$ 1.94  & -10.83 $\pm$ 0.67 & 49.64 & -82.23 & -28.43 & $-1.379^{+0.002}_{0.002}$ & 3100   & $3.8^{+0.74}_{-0.8}$    & $27.81^{+11.47}_{-7.89}$ \\
UCAC4 059-012851             & 101.641 & ...                & ...                & ...               & 50.26 & -83.98 & -27.42 & $-1.222^{+0.015}_{0.016}$ & 2900   & ...                     & $4.01^{+1}_{-1.54}$      \\
eps cha 12                   & 106.723 & -11.17 $\pm$ 1.4   & -17.79 $\pm$ 2.37  & -9.6 $\pm$ 0.77   & 52.39 & -88.44 & -28.69 & $-1.425^{+0.002}_{0.003}$ & 3200   & $6.11^{+2.94}$          & $5.02^{+3.49}$           \\
Gaia DR3 5837882207728535040 & 104.036 & ...                & ...                & ...               & 50.65 & -87.6  & -24.18 & $-2.548^{+0.003}_{0.003}$ & 2600   & $11.09^{+5.66}_{-1.09}$ & $1.02^{+0.45}$           \\
UCAC4 105-048794             & 96.905  & ...                & ...                & ...               & 46.1  & -84.5  & -11.18 & $-1.183^{+0.003}_{0.002}$ & 3400   & $10.69^{+6.14}_{-1.8}$  & $5.81^{+1.48}_{-1.09}$   \\
2MASS J12003792-7845082      & 101.641 & ...                & ...                & ...               & 49.37 & -84.24 & -28.24 & $-1.94^{+0.002}_{0.003}$  & 3000   & $10.02^{+2.12}_{-2.42}$ & ...                      \\
UCAC4 056-012157             & 101.854 & ...                & ...                & ...               & 49.43 & -84.4  & -28.41 & $-1.081^{+0.002}_{0.002}$ & 3100   & $2^{+0.56}_{-0.81}$     & $8.92^{+2.44}$           \\
Gaia DR3 5227167240115285248 & 101.48  & -11.62 $\pm$ 2.17  & -17.52 $\pm$ 3.91  & -8.44 $\pm$ 0.8   & 48.47 & -87.34 & -17.91 & $-1.592^{+0.021}_{0.022}$ & 2900   & $4^{+0.02}_{-1.97}$     & $10^{+2.57}_{-1.98}$     \\
Gaia DR3 5200024318108381568 & 102.95  & -5.41 $\pm$ 8.74   & -27.9 $\pm$ 15.09  & -13.92 $\pm$ 5.16 & 49.5  & -85.41 & -29.22 & $-0.972^{+0.008}_{0.008}$ & 3000   & ...                     & $36.87^{+7.82}_{-6.39}$  \\
UCAC4 096-034374             & 102.681 & -10.02 $\pm$ 10.64 & -18.03 $\pm$ 19.9  & -7.25 $\pm$ 3.44  & 47.87 & -89.52 & -15.46 & $-0.953^{+0.012}_{0.012}$ & 3300   & $3.3^{+1}_{-0.53}$      & $3.55^{+2.08}_{-1.63}$   \\
Gaia DR3 5233289252143166336 & 104.446 & ...                & ...                & ...               & 48.75 & -90.84 & -16.74 & $-2.278^{+0.003}_{0.002}$ & 2600   & $3.46^{+3.44}_{-1.81}$  & $12.65^{+4.88}_{-2.84}$  \\
Gaia DR3 5227282517035989632 & 98.469  & -8.14 $\pm$ 2.01   & -19.81 $\pm$ 3.8   & -9.02 $\pm$ 0.81  & 45.22 & -85.57 & -18.15 & $-1.737^{+0.002}_{0.002}$ & 2900   & $4.03^{+1.09}_{-0.6}$   & $11.46^{+4.39}_{-2.64}$  \\
UCAC4 060-011194             & 100.295 & -10.79 $\pm$ 2     & -19.64 $\pm$ 3.64  & -10.22 $\pm$ 1.17 & 46.55 & -84.56 & -27.22 & $-0.996^{+0.003}_{0.002}$ & 3400   & $6.19^{+2.13}_{-1.28}$  & $12.59^{+4.38}$          \\
UCAC4 086-024494             & 101.229 & -7.7 $\pm$ 1.94    & -24.28 $\pm$ 4.04  & -10.63 $\pm$ 0.91 & 42.89 & -89.45 & -20.17 & $-1.254^{+0.016}_{0.016}$ & 3000   & $2^{+0.41}_{-0.14}$     & $38^{+0.59}$             \\
UCAC4 101-037476             & 99.447  & ...                & ...                & ...               & 39.79 & -89.84 & -15.31 & $-1.178^{+0.02}_{0.02}$   & 3300   & $6.11^{+3.07}_{-1.48}$  & $8.69^{+2.51}_{-2.11}$   \\
2MASS J10563146-7618334      & 99.66   & -10.47 $\pm$ 1.46  & -19.7 $\pm$ 2.97   & -9.86 $\pm$ 0.88  & 42.46 & -86.41 & -25.74 & $-1.285^{+0.015}_{0.015}$ & 3200   & $4.89^{+1.79}_{-0.95}$  & $39.64^{+5.36}$          \\
Gaia DR3 5199332484776169600 & 95.446  & ...                & ...                & ...               & 40.57 & -81.4  & -28.95 & $-1.881^{+0.002}_{0.002}$ & 2900   & $5.08^{+1.17}_{-0.08}$  & $3.56^{+0.86}_{-0.5}$    \\
Gaia DR3 5229663337667889152 & 100.058 & ...                & ...                & ...               & 37.52 & -90.5  & -20.34 & $-2.254^{+0.003}_{0.002}$ & 2900   & $18.32^{+3.3}_{-6.47}$  & $14.27^{+3.49}_{-2.24}$
\enddata

\end{deluxetable*}
\end{longrotatetable}
\subsection{Age Estimates}

DV+21 used a single star locus fit to derive the age of the ECA, establishing an age range of 3--8 Myr (i.e., $5^{+3}_{-2} $ Myr). This range encompasses estimates found by \citet{murphy_re-examining_2013} as well as from the younger LCC members compiled in \citet{goldman_large_2018}. \citet{goldman_large_2018} determined their ages from two different isochrone grids, i.e., the PADOVA Parsec 3.0 grids \citep{bressan_span_2012} for high mass stars and the BT-settl grid for low mass stars. For the A0 LCC subgroup, \citet{goldman_large_2018} determined an average age of 7.0 Myr.

We determine the ages of individual stars in our sample of new candidates, as we well as previously identified ECA and LCC A0 stars, using VOSA's stellar age estimator\footnote{\url{http://svo2.cab.inta-csic.es/theory/vosa/}}. This tool uses the luminosity and effective temperature, as determined from SED fitting (Fig.~\ref{fig:sed fit}), and obtains an estimated age by interpolating between the nearest two isochrones in a given model isochrone grid that lie on either side of the star's HR diagram position. For our age estimates, we chose the BT-SETTL and SPOTS isochrone grids. We apply SPOTS models with solar metallicity and filling factor $f = 0.34$, where $f$ is defined as the percentage of the stellar surface that is covered by starspots \citep{somers_spots_2020}.

\begin{figure}
    \centering
    \includegraphics[width = 1.0\textwidth]{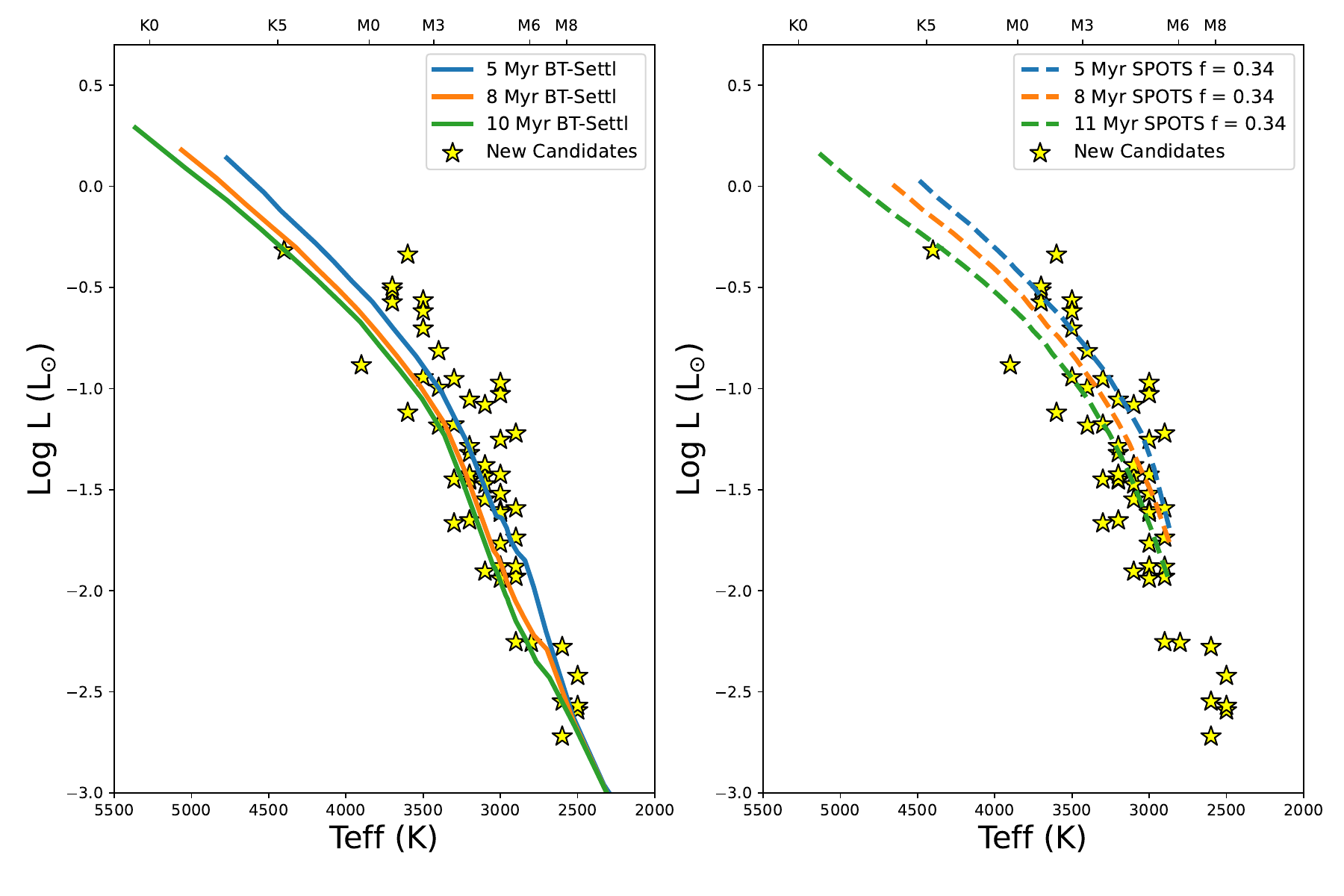}
    \caption{HR diagram positions of our new ECA candidates (yellow stars) overlaid with BT-SETTL model isochrones for 5, 8, and 10 Myr (left) and SPOTS model isochrones, with a spot filling factor of f = 0.34, for 5, 8, and 11 Myr (right).}
    \label{fig:isochrone comparison}
\end{figure}

\begin{figure}
    \centering
    \includegraphics[width = 1.0\textwidth]{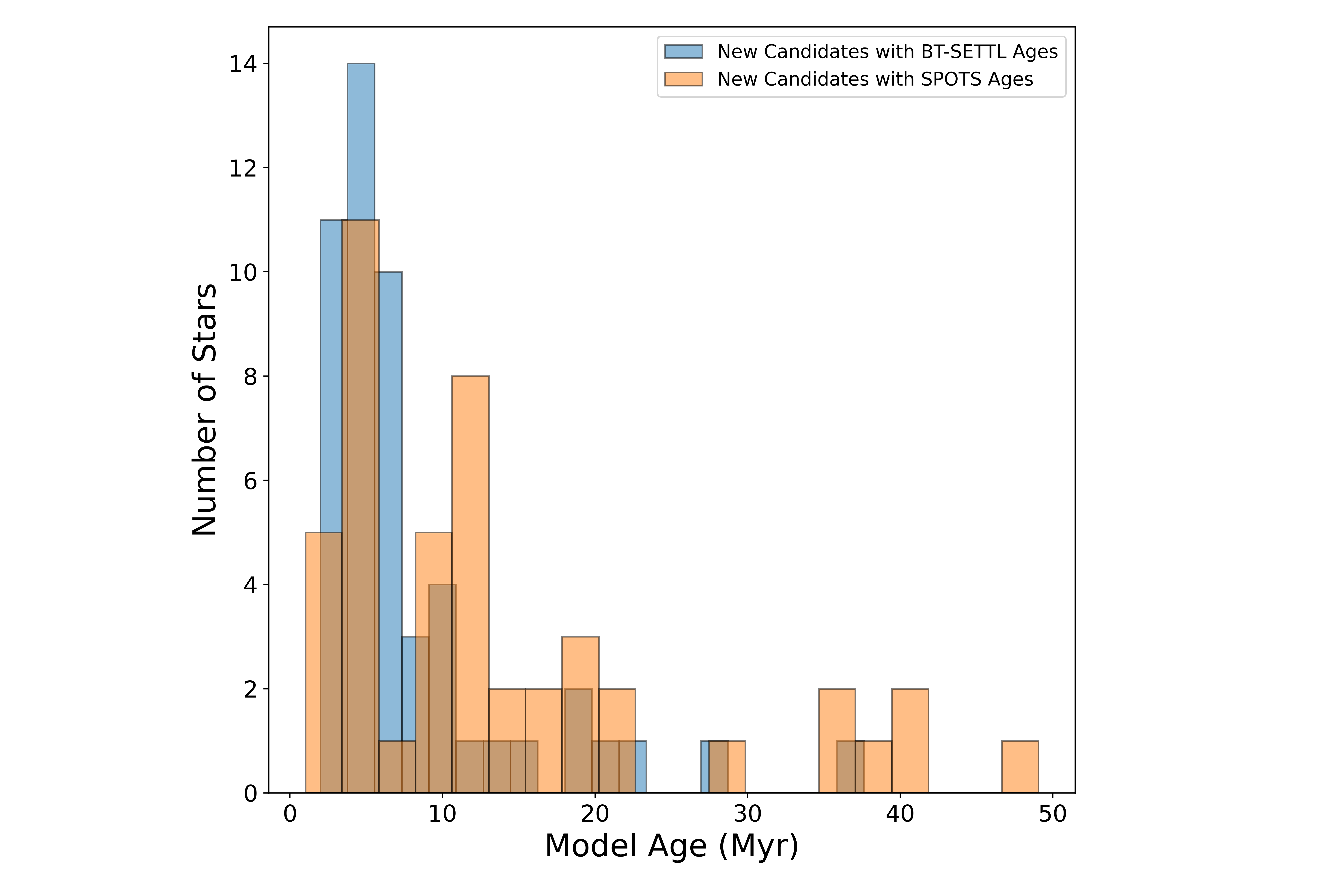}
    \caption{Age distribution for the new ECA candidates. Blue bars shows stellar ages obtained with the BT-SETTL model. Orange bars show ages obtained from the SPOTS model adopting a filling factor of f = 0.34.}
    \label{fig:age distribution}
\end{figure}

In Fig. \ref{fig:isochrone comparison} we present stellar isochrones from these two models with our new ECA candidates overplotted.  This plot highlights how these models diverge in terms of age estimates for mid- to late-M dwarfs. For young M dwarfs later than M3, The isochrones from the SPOTS model with a filling factor $f = 0.34$ yield larger luminosity estimates than those yielded by the BT-SETTL models, resulting in older age estimates from the SPOTS models.
In Fig. \ref{fig:age distribution}, we show the distribution of estimated stellar ages from both models, illustrating this systematic effect. Application of the SPOTS models results in increased average ages for both the LCC A0 and ECA groups.

\begin{figure}

  \centering
  \includegraphics[width=0.8\linewidth]{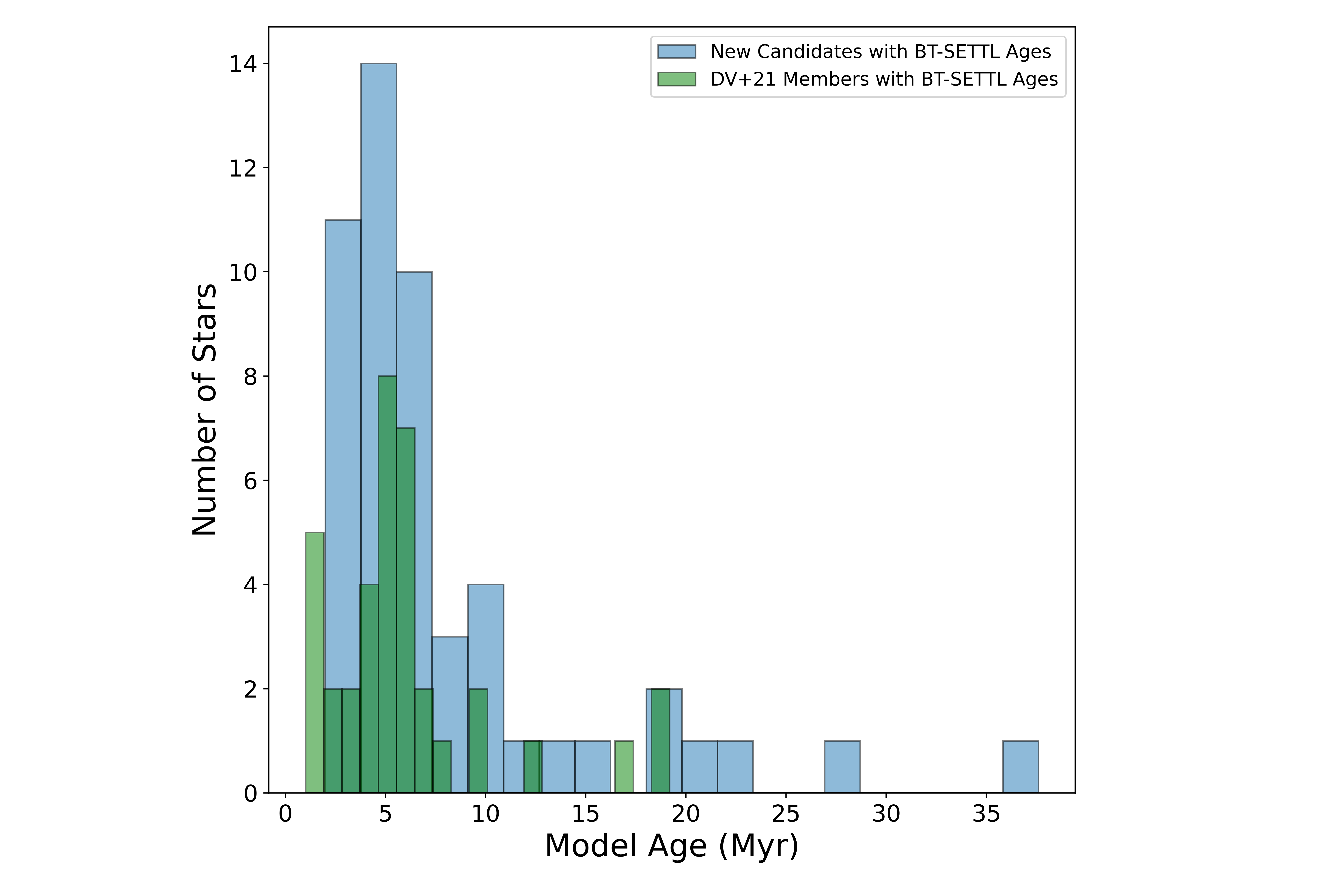}
  \centering
  \includegraphics[width=0.8\linewidth]{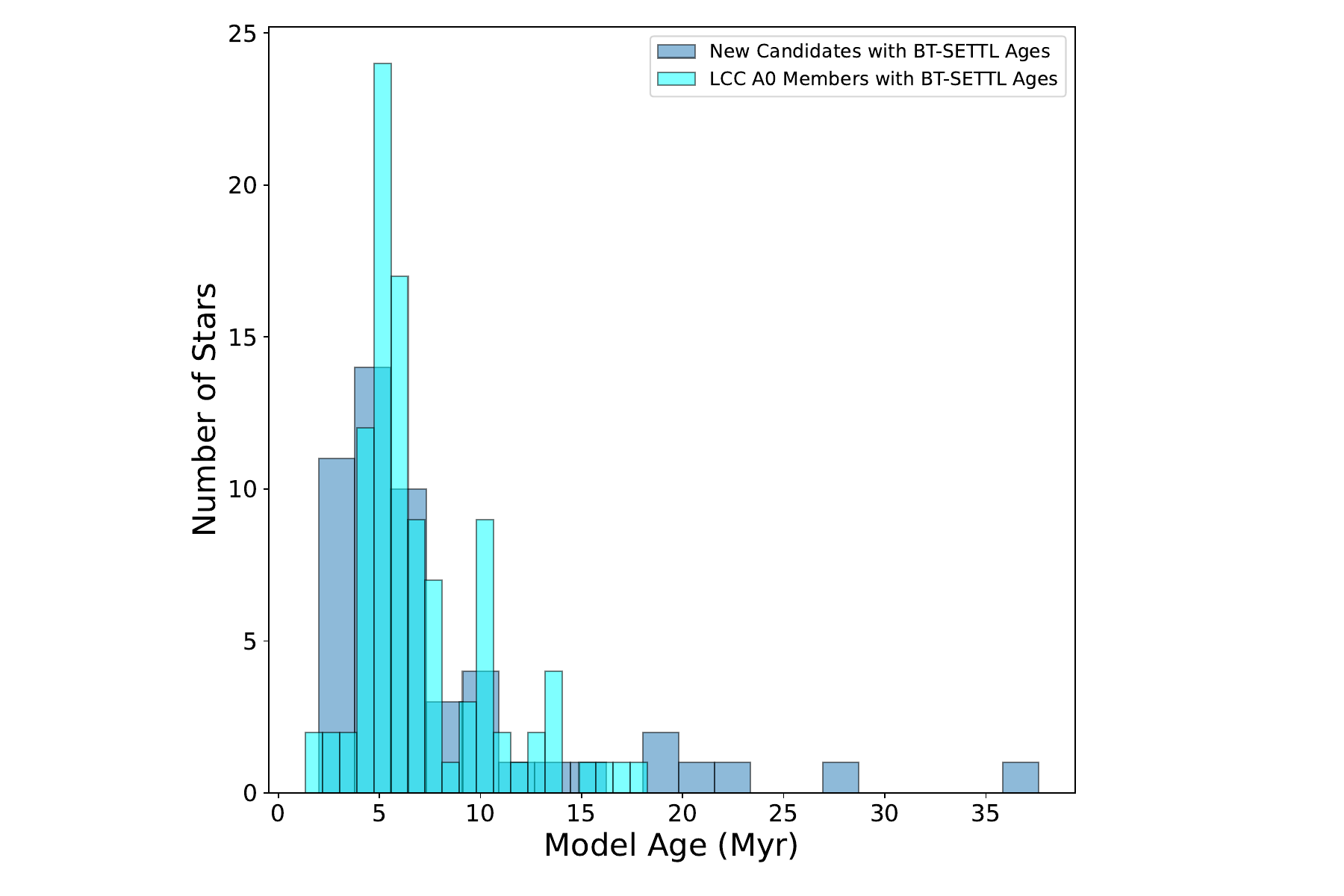}

\caption{Comparisons of age distributions of new ECA candidates (dark blue) with previously identified ECA members (green, top plot) and with LCC A0 stars (teal; bottom plot). All ages are derived using BT-SETTL model isochrones.}
\label{fig:twoagecomp}
\end{figure}

We also estimate ages for both the ECA members from DV+21 and LCC members from G+18 using the BT-Settl model, so as to have a consistent method for age estimation and comparison to our new candidates. 
In Fig. \ref{fig:twoagecomp} we compare each group's BT-Settl-based age distribution to that of our new candidates. While the age distribution of our new candidate sample closely aligns with that of the ECA, the LCC A0 group age distribution includes nearly twice as many young stars in the 5--7 Myr range, and exhibits a second peak at 10 Myr. 
The mean and median ages we derive for the ECA, LCC A0, and our new candidates for the ECA are listed in Table~\ref{table:meanage}.

\begin{deluxetable*}{lCCC}
\tablecaption{\sc BT-SETTL-derived mean and median ages of ECA and LCC candidates \label{table:meanage}}
\tablewidth{0pt}
\tablehead{
\colhead{Group} & \colhead{No.} & \colhead{Mean age}  & \colhead{Median age} \\
\colhead{}&\colhead{} & \colhead{(Myr)} & \colhead{(Myr)} 
}
\startdata
Established ECA Candidates (DV+21)&50 & 6.07$\pm$0.620 & 5.17 \\
New ECA Candidates (this work) & 54 & 8.01$\pm$0.937  & 5.82 \\
LCC A0 (G+18) &111& 7.05$\pm$0.314  & 5.99 \\
\enddata
\end{deluxetable*}

\subsection{IR Excesses}\label{sec:IRexcesses}

We have investigated which of our new candidates display evidence for an infrared excess above their stellar photospheric emission. The presence of such an IR excess serves as evidence of warm dust that is likely indicative of the presence of a circumstellar disk. \citet{pecaut_star_2016} and \citet{luhman_disk_2012} proposed a classification method based on 2MASS-WISE IR color excesses $E(K_s - W3)$ and $E(K_s - W4)$ for disk classes as detailed in \citet{espaillat_transitional_2012}. In this system, stars' IR excesses are assessed in terms of classifications as likely debris (gas-poor) disks, transitional disks (i.e., disk with central cavities), full disks (likely gas-rich and lacking central cavities), and evolved disks (i.e., disks in an intermediate state between debris and transition/full disks).

We used photometric data from 2MASS and WISE and corresponding model colors from \citet{pecaut_intrinsic_2013} to identify those new candidates that display infrared excesses $E(K_s - W3)$ and $E(K_s - W4)$. We examined each candidate's SED by eye to identify potentially spurious WISE W3 and W4 band photometry. These bands are most sensitive to background extragalactic sources as well as intervening or background ISM dust emission. 
We also examined each WISE postcard image \citep{wright_wide-field_2010} to assess the possibility of source contamination as well as whether the observations are consistent with excess emission due to a disk. 

\begin{figure}
    \centering
    \includegraphics[width = 1.0\textwidth]{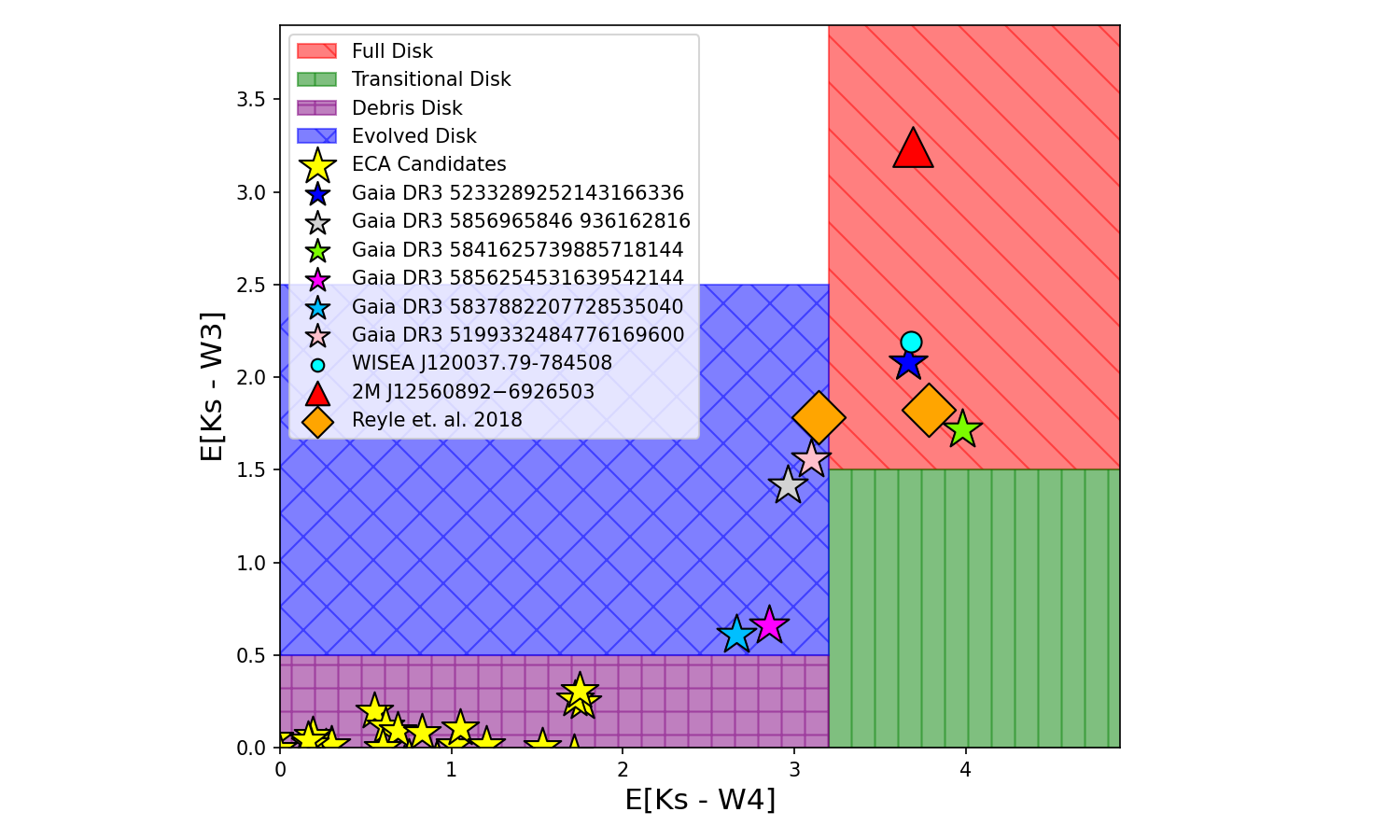}
    \caption{Infrared excesses, $E(K_s - W3)$ vs.\ $E(K_s - W4)$, for all new candidate stars; stars with the largest excesses are highlighted (with symbols as indicated in the legend). Background colors indicate IR-excess-based disk classifications as detailed in \citet{pecaut_star_2016}. 
    }
    \label{fig:IR excess}
\end{figure}

We find that most of our candidates do indeed display possible or definitive WISE excesses, i.e., they display values of both $E(Ks - W3)$ and $E(Ks - W4)$ $> 0$.
Fig. ~\ref{fig:IR excess} displays inferred infrared excesses $E(K_s - W3)$ and $E(K_s - W4)$ for our candidates, overlaid on the \citet{pecaut_star_2016} IR-excess-based disk classification scheme. We expect that most ECA and LCC stars with significant excesses indicative of full, evolved, or transition disks to already have been identified prior to our study \citep{goldman_large_2018,dickson-vandervelde_gaia-based_2021,kubiak_new_2021}. Indeed, most of our candidates lie in the range $1 > E(Ks - W3) > 0$, indicative of debris disks. However there are ten or so stars with larger excesses that warrant followup observations. The bottom row of panels in Fig.~\ref{fig:sed fit} shows four stars among our candidates that show such significant infrared excesses.

\subsubsection{Notable IR Excess Identified Candidates }
We have identified eight very cool, low-mass candidates with significant IR excesses, six of which are without previous literature references. Photometric data and best-fit effective temperatures for these six candidates are listed in Table \ref{tab:dwarfs}. These stars all lie near or below the $\sim$M6 spectral type boundary between very low-mass M dwarfs and future/eventual brown dwarfs of age $\sim$5--10 Myr \citep{baraffe_new_2015} HERE NEW paper.  This indicates that some or all of these objects are proto-brown dwarfs. Spectroscopic observations to confirm our spectral type assessments of these objects are hence warranted.

{\it{WISEA J120037.79$-$784508:}}
We have recovered the object WISEA J120037.79-784508 among our candidate ECA members. First identified by \citet{schutte_discovery_2020}, this star was determined to be a disk-bearing young brown dwarf. Indeed, in our parameter estimation (Table \ref{table:estimated params}), we also find that this object lies on the M dwarf/brown dwarf boundary and has a significant IR excess. \citet{schutte_discovery_2020} also determined that this object has a high probability to be a member of the ECA. Our modeling supports this assessment and further bolsters our method to recover potential new members of the ECA among our sample.

{\it{2MASS J12560892-6926503:}}
The 2MASS source J12560892-6926503 appears to have among the strongest IR excesses among our candidates (Fig.~\ref{fig:IR excess}). However, its apparent IR excess is likely a spurious result due to binarity-related source confusion. This object was previously identified in a search for nearby young wide separation binaries \citep{bohn_unveiling_2022}. As part of our recovery of nearby stars around the ECA, we have also identified a likely companion to 2MASS J12560892-6926503, 1RXS J125608.8-692652. The stellar counterpart to this ROSAT source was first identified as a Sco-Cen candidate in \citet{song_new_2012}.
In analyzing the SED for 2MASS J12560892-6926503, we concluded that its unusually large excess is most likely due to contamination from this likely companion. In particular, the J band flux appears to be associated with 1RXS J125608.8-692652. However, 2MASS J12560892-6926503 and the J-band 2MASS counterpart to 1RXS J125608.8-692652 appear to be resolved by Gaia, with distinct Gaia positions and photometry. This photometry was used to fit a model to the SED of 2MASS J12560892-6926503. Its spectral type is estimated to be M6, while we find the optical/IR counterpart to 1RXS J125608.8-692652 has a spectral type of M3.

\begin{deluxetable*}{lcccccccc}
\tablecaption{\sc Newly identified cool dwarfs with significant IR excesses\label{tab:dwarfs}}
\tablewidth{0pt}
\tablehead{
\colhead{Name} & \colhead{$G$} & \colhead{$K_s$} & \colhead{$W3$} & \colhead{$W4$} & \colhead{$T_{eff}$}\\
\colhead{}  & \colhead{(mag)} & \colhead{(mag)} & \colhead{(mag)} & \colhead{(mag)} & \colhead{(K)}}
\startdata
Gaia DR3 5233289252143166336 & 12.5182 & 12.379 & 9.675  & 7.963 & 2600 \\
Gaia DR3 5856965846936162816 & 10.0891  & 10.699 & 8.74   & 7.07  & 3100 \\
Gaia DR3 5841625739885718144 & 11.6371 & 10.669 & 8.33   & 5.939 & 3100 \\
Gaia DR3 5856254531639542144 & 13.4260 & 13.173 & 11.812 & 9.485 & 2500 \\
Gaia DR3 5837882207728535040 & 13.3564 & 13.119 & 11.805 & 9.622 & 2600 \\
Gaia DR3 5199332484776169600 & 11.0866 & 11.328
 & 9.225  & 7.56 & 2900
\enddata

\end{deluxetable*}

\section{Discussion}

\subsection{The (vanishing) boundary between the ECA and LCC}

Over the last couple decades, the number of known young moving groups within 200 pc has grown from a half-dozen or so \citep[e.g.,][]{torres_young_2008} to more than 30 individual groups \citep{gagne_number_2021}. This rapid growth is in part the result of a multifaceted approach to identifying young stars \citep{gagne_banyan_2017,kerr_stars_2021,higashio_disks_2022}. With the advent of the Gaia mission, identification of nearby, young stars has only become easier. Yet, as Gaia approaches a near-complete census of all stars with $G < 14$ within $\sim$200 pc, the Galactic spatio-kinematic distributions of individual moving groups have become more ambiguous. Gaia is gradually replacing the original picture of nearby young moving groups as comoving, coeval loose associations with memberships of  $< 100$ stars, revealing larger interconnected structures with groups approaching $\sim 1000$ members and potentially sharing common star formation histories \citep{zari_3d_2018,kounkel_untangling_2019,kerr_stars_2021,gagne_number_2021}. In most cases, membership is established based on clear kinematic differences between groups. However, stars lying on the spatial borders and overlap regions between groups may also share similar kinematics, complicating the process. For example, the Tuc and Hor moving groups were originally identified as separate entities, until later studies demonstated that they belong in a unified group, renamed as simply Tuc-Hor \citep{zuckerman_tucanahorologium_2011}. Given its similar age and kinematics, the Columba Association may be an additional extension of Tuc-Hor \citep{zuckerman_nearby_2019}. 

The LCC and ECA were originally considered to be a single group \citep{mamajek_post-t_2002,torres_young_2008,pecaut_star_2016}; however, more recent analyses indicated they may be distinct in terms of both spatio-kinematics and age \citep{murphy_re-examining_2013,goldman_large_2018}.  In our analysis of the positions of existing ECA and LCC members, 10 out of 54 candidates lie within the `core' ECA region. However, the remaining candidates reside within the region bounded by the lower end of the LCC (see Fig.~\ref{fig:position_comparison}). Furthermore, ECA members identified in DV+21 but not considered in \citet{murphy_re-examining_2013} also occupy this region. These new, Gaia-based ECA candidate identifications hence appear to blur the spatial lines between the ECA and LCC. 

The clustering in velocity space (Fig. \ref{fig:tangv}, \ref{fig:velocity_comparison}) as well as the average velocity components for our new candidates, ECA members, and the LCC A0 (Table~\ref{table:motion}) also indicates more similarities than differences between the two groups. The older subgroups of the LCC also appear to show a deviation in terms of their average velocities (see Table 1 in \citep{goldman_large_2018}), indicating a correlation with group dispersion and age. 

When considering inferred ages, Fig.~\ref{fig:twoagecomp} and Table~\ref{table:meanage} demonstrate that the stellar age distributions of the new candidates, DV+21 ECA members, and G+18 A0 members do not diverge significantly. The southern end of the LCC does appear to include a mix of older stars \citep{goldman_large_2018}, which is reflected in the A0 group age histogram. 
This however does not meaningfully change the average age of the A0 subgroup or even the general clustering towards the LCC/ECA boundary. In their discussions, both \citet{murphy_re-examining_2013} and \citet{kubiak_new_2021} consider that there may be no clear boundary between the LCC and ECA. Indeed, in a global analysis considering all known moving groups, \citet{kerr_stars_2021} found that the ECA can be considered a subgroup of LCC, denoting the ECA as LCC A. 

\subsection{Circumstellar disks in the ECA/LCC boundary region}

Stars in the age range $\sim$5--10 Myr are more likely to exhibit debris disks than ``full'' (gaseous) or transition disks. This is reflected in our analysis of new ECA candidates:
only 10 out of 54 stars among our new ECA candidates in the ECA/LCC boundary region show IR excesses indicative of full, transitional, or evolved disks (Fig.~\ref{fig:IR excess}). Included among these newly identified disk-bearing stars are six ultra-low-mass, mid- to late-M stars that lie near the future hydrogen-burning limit. 

In the context of the broader ECA population, this brings the overall disk fraction to $20\%$ excluding debris disks, and $40\%$ including debris disks. Based on DR2 data, DV+21 found a disk fraction of $30\%$. Including the previously identified disks, $80\%$ of all disk-hosting stars in the ECA are of spectral type M. This is consistent with previous observations indicating that low-mass, late-M-type stars tend to retain their disks over $\sim$ 10 Myr timescales \citep{kastner_m_2016,silverberg_peter_2020,gaidos_planetesimals_2022}. \citet{goldman_large_2018} found evidence for disks for 115 of 1002 stars across all spectral types in their age 7--20 Myr sample, for a disk fraction of $10\%$. These disk fractions for the LCC and ECA generally follow disk dissipation relations determined by \citet{michel_bridging_2021}.

\section{Summary and Conclusions}
At a distance of just $\sim$100 pc and age range of just $\sim$5--10 Myr, the population spanning the boundary between the ECA and LCC provides a unique site for the study of low-mass, pre-main sequence stellar evolution. In our search of new members of the ECA using the Gaia DR3 data release, we identify 54 new candidates. Combined with past studies, this would bring the total membership of the ECA to $\sim$90 systems. 

However, our new ECA candidates also appear to further blur the distinction between the youngest LCC stars and the ECA. New candidates that we have selected are clearly isochronally young, consistent with the young ($\sim$5--10 Myr) ages of both the ECA and the G+18 LCC A0 subgroup. Through analysis of heliocentric positions and kinematics, we arrive at the conclusion that the ECA is sufficiently kinematically similar to the youngest stars in the LCC to be considered a subgroup of that (generally older)  association. However, there exists a denser core ECA region that could stand on its own as a distinct and unique group of younger low-mass, pre-MS stars. Our SED fitting and analysis, which provides new age estimates for stars in the ECA and the LCC A0 group, supports a mean age of $\sim$5 Myr for the stars in the ECA core; we find ECA stars overlapping with the youngest LCC members have a mean age closer to $\sim$8 Myr. One interpretation of this result is that the ECA and LCC formed within the same molecular cloud, with star formation initiating in the LCC and propagating to the ECA. This would explain the age gradient between the subgroups.

The fraction of new candidates with disks is found to be $20\%$. When including debris disks the fraction increases to $50\%$. If added to the established ECA stars, the disk fraction remains at $20\%$ but the fraction including debris disks drops to $40\%$.
We also report the recovery among our ECA candidates of six previously unidentified ultra-low-mass, mid- to late-M stars, lying near the future hydrogen-burning limit, that display significant infrared excesses. Given that the young late-type-M/brown-dwarf boundary is largely unexplored for stars in the $\sim$5--10 Myr age range of the ECA and LCC A0 group, followup observations that can better establish the natures of these objects are warranted.

\section{Acknowledgements}
This research was supported by NASA Astrophysics Data Analysis Program grant 80NSSC22K0625 and NASA Exoplanets Program grant 80NSSC19K0292 to RIT. This work has made use of data from the European Space Agency (ESA) mission {\it Gaia} (\url{https://www.cosmos.esa.int/gaia}), processed by the {\it Gaia} Data Processing and Analysis Consortium (DPAC,
\url{https://www.cosmos.esa.int/web/gaia/dpac/consortium}). Funding for the DPAC has been provided by national institutions, in particular the institutions participating in the {\it Gaia} Multilateral Agreement.
This publication makes use of VOSA, developed under the Spanish Virtual Observatory (https://svo.cab.inta-csic.es) project funded by MCIN/AEI/10.13039/501100011033/ through grant PID2020-112949GB-I00.
VOSA has been partially updated by using funding from the European Union's Horizon 2020 Research and Innovation Programme, under Grant Agreement nº 776403 (EXOPLANETS-A)
This publication makes use of data products from the Two Micron All Sky Survey, which is a joint project of the University of Massachusetts and the Infrared Processing and Analysis Center/California Institute of Technology, funded by the National Aeronautics and Space Administration and the National Science Foundation.
This publication makes use of data products from the Wide-field Infrared Survey Explorer, which is a joint project of the University of California, Los Angeles, and the Jet Propulsion Laboratory/California Institute of Technology, funded by the National Aeronautics and Space Administration.
This research has made use of the SIMBAD database,
operated at CDS, Strasbourg, France.
This work made use of Astropy:\footnote{http://www.astropy.org} a community-developed core Python package and an ecosystem of tools and resources for astronomy\citep{collaboration_astropy_2022}.

\bibliography{EpsChaPaperText}{}
\bibliographystyle{aasjournal}

\end{document}